\documentclass[preprint,onecolumn,nofootinbib]{revtex4}
\pdfoutput=1
\usepackage[colorlinks=true,linkcolor=blue,urlcolor=blue,filecolor=black,citecolor=red,pdfstartview=FitV,pdftitle={},pdfsubject={},pdfkeywords={},pdfpagemode=None,bookmarksopen=true]{hyperref}
\usepackage{graphicx}
\usepackage{amsmath}
\usepackage{amsfonts}
\usepackage{amssymb,ulem}
\usepackage{color,xcolor}%
\usepackage{CJK}
\usepackage{subfigure}
\usepackage{float}
\usepackage{ulem}

\usepackage{dcolumn}
\usepackage{verbatim}
\usepackage{graphicx}
\usepackage{cancel}
\usepackage{booktabs}  
\usepackage{multirow}  

\begin{document}
\title{Quasinormal modes of a charged loop quantum black hole}

\author{Li-Gang Zhu$^{1}$}
\thanks{zlgoupao@163.com}
\author{Guoyang Fu$^{1}$}
\thanks{FuguoyangEDU@163.com}
\author{Shulan Li$^{2}$}
\thanks{shulanli.yzu@gmail.com}
\author{Dan Zhang$^{1}$}
\thanks{danzhanglnk@163.com}
\author{Jian-Pin Wu$^{1}$}
\thanks{jianpinwu@yzu.edu.cn} 
\affiliation{
  $^1$ Center for Gravitation and Cosmology, College of Physical
  Science and Technology, Yangzhou University, Yangzhou 225009,
  China\\
  $^2$ Department of Physics, Shanghai University, Shanghai 200444, China}

\begin{abstract}

This study presents a systematic investigation of quasinormal modes (QNMs) for probe fields—massless/massive scalar and Dirac fields—around a charged loop quantum gravity black hole (LQG-BH) characterized by the quantum parameter $b_0$ and the charge parameter $Q$. Through spectral analysis of quasinormal frequencies (QNFs), we uncover a distinct overtone outburst driven by quantum gravity effects, prominently manifested in the scalar field spectrum with the multipole quantum number $l=0$. Both the outburst and its accompanying oscillatory patterns grow more pronounced with increasing overtone numbers. In contrast, massless scalar fields with $l>1$ and Dirac fields exhibit delayed outburst development, with non-monotonic behavior dominating the first two overtones. Notably, increasing the charge $Q$ universally suppresses quantum-gravity-induced features, including outbursts, non-monotonicity, and oscillations. Furthermore, we present evidence suggesting the presence of quasi-resonances in the massive scalar QNM spectrum, thereby illustrating the potential for the emergence of arbitrarily long-lived modes in this charged LQG spacetime. These findings establish a robust and universal interplay between quantum gravity effects and charge dynamics, providing new insights into the spectral properties of quantum-corrected BHs.

\end{abstract}

\maketitle
\tableofcontents

\section{Introduction}

Loop quantum gravity (LQG) is considered a promising contender for the theory of quantum gravity, characterized by its non-perturbative and background-independent nature \cite{Rov,Thiemann:2001gmi,Ashtekar:2004eh,Han:2005km}. The quantization technique developed in full LQG has been effectively employed in the symmetry reduced cosmological model, resulting in the development of loop quantum cosmology (LQC) \cite{Bojowald:2001xe,Ashtekar:2006rx,Ashtekar:2006uz,Ashtekar:2006wn,Ashtekar:2003hd,Bojowald:2005epg,Ashtekar:2011ni,Wilson-Ewing:2016yan}. An outstanding characteristic of LQC is its ability to naturally substitute the classical big bang singularity of the universe with a quantum bounce \cite{Bojowald:2001xe,Ashtekar:2006rx,Ashtekar:2006uz,Ashtekar:2006wn,Ashtekar:2003hd,Bojowald:2005epg,Ashtekar:2011ni,Wilson-Ewing:2016yan,Bojowald:2003xf,Singh:2003au,Vereshchagin:2004uc,Date:2005nn,Date:2004fj,Goswami:2005fu,Papanikolaou:2023crz}, leading to a non-singular evolution of the universe \cite{Bojowald:2005zk,Stachowiak:2006uh}.

The approach developed in LQC can be readily applied to the spherically symmetric Schwarzschild black hole (BH) model, thereby opening up the field of loop quantum gravity black hole (LQG-BH). For a detailed construction of the model, please refer to \cite{Ashtekar:2005qt,Boehmer:2007ket,Chiou:2008nm} and also see the reviews \cite{Perez:2017cmj,Zhang:2023yps}. Just as LQC can address the Big Bang singularity of the universe, the LQG-BH model can likewise resolve the interior singularity of BHs. Specifically, for the majority of LQG-BHs, the singularity is substituted with a transition surface that connects a trapped region and an anti-trapped region. 

Contrary to the LQC, which usually has a consistent treatment for various models, within the framework of the LQG-BH model, there are various models that employ different schemes to regularize and quantize the Hamiltonian constraint. In general, the LQG-BH models can be categorized into three main schemes: the $\mu_0$-scheme, the $\bar{\mu}$-scheme, and the generalized $\mu_0$-scheme. In the $\mu_0$-scheme, it is assumed that the quantum regularization parameters remain constant over the whole phase space \cite{Ashtekar:2005qt,Modesto:2005zm,Modesto:2008im,Campiglia:2007pr,Bojowald:2016itl}. An inherent drawback of this approach is that the final outcome is dependent on the fiducial structures introduced in the construction of the classical phase space. In addition, even in situations with low curvature, notable quantum effects may emerge, rendering these models non-physical. To eliminate the dependency on fiducial structures, the $\bar{\mu}$-scheme is proposed, in which the quantum regularization parameters are selected as a function of the phase space variables \cite{Boehmer:2007ket,Joe:2014tca,Chiou:2008nm,Chiou:2008eg}. In particular, in this scheme with Choui's choice \cite{Chiou:2008nm,Chiou:2008eg}, the spacetime rapidly converges to the Schwarzschild geometry when the curvatures are low, which addresses the disadvantage of the $\mu_0$-scheme \cite{Gambini:2020nsf,Kelly:2020uwj,Husain:2022gwp,Han:2022rsx}. Nevertheless, the $\bar{\mu}$-scheme is also subject to the drawback that the quantum corrections to the Schwarzschild BH horizon may be substantial, contrary to the prevailing belief that the horizon is a classical region and should not experience significant quantum corrections \cite{Boehmer:2007ket,Chiou:2008nm}. To alleviate the aforementioned issues, several authors have recently proposed the generalized $\mu_0$-scheme \cite{Corichi:2015xia,Olmedo:2017lvt,Ashtekar:2018lag,Ashtekar:2018cay}. In addition, the quantum collapsing model introduced in \cite{Lewandowski:2022zce} offers an additional approach to mitigating these problems.

Quantum gravity effects inevitably induce subtle modifications to the effective background spacetime geometry, consequently manifesting as discernible signatures in the quasinormal mode (QNM) spectra.
Particularly, a remarkable advancement has been the establishment of a connection between overtones and near-horizon geometry, first reported in Ref. \cite{Konoplya:2022pbc}. This study found that even a slight near-horizon deformation can result in a sudden and significant change in the overtones, a phenomenon termed the ``outburst of overtones” and metaphorically described as the ``sound of the event horizon”. Subsequently, numerous studies on overtones have further demonstrated and confirmed this observation \cite{Berti:2005ys,Berti:2018vdi,Fu:2022cul,Fu:2023drp,Gong:2023ghh,Moura:2021eln,Moura:2021nuh,Moura:2022gqm,Lin:2024ubg,Ghosh:2022gka,Konoplya:2023aph,Zinhailo:2023xdz,Zhang:2024nny,Song:2024kkx,Konoplya:2024lch,Stashko:2024wuq,Dubinsky:2024nzo,Zinhailo:2024kbq,Dong:2024ams,Livine:2024bvo}.
Moreover, the studies \cite{Fu:2022cul,Fu:2023drp,Gong:2023ghh,Zhang:2024nny,Song:2024kkx} also indicate that in some effective quantum gravity model, a non-monotonic behavior of the quasinormal frequencies (QNFs) as a function of the corrected parameters is observed as the extremal BH is approached. This phenomenon manifests most prominently in the fundamental mode characterized by the lowest multipole quantum number. As the multipole quantum number increases, the non-monotonic behavior becomes less pronounced and gradually disappears. This suggests that the influence of the multipole quantum number begins to dominate over the effects of quantum gravity \cite{Song:2024kkx,Fu:2022cul,Fu:2023drp,Gong:2023ghh,Zhang:2024nny}. However, we emphasize that for higher multipole quantum number, the non-monotonic behavior reemerges at higher overtones \cite{Zinhailo:2024kbq,Konoplya:2022hll,Bolokhov:2023bwm,Fu:2022cul,Gong:2023ghh,Zhang:2024nny,Dong:2024ams}, albeit developing more slowly due to the dominance of the centrifugal barrier in the effective potential.
Another interesting observation is that an oscillatory pattern in the overtones as a function of the corrected parameter is detected as the extremal BH is approached \cite{Fu:2022cul,Fu:2023drp,Gong:2023ghh,Zhang:2024nny}. Both the non-monotonic behavior in the fundamental modes and the oscillatory pattern in the overtones may be linked to the extremal BH.

Recently, a novel and intriguing uniparametric polymerisation scheme\footnote{The phase space regularization technique employed in LQG is also known as polymerization \cite{Corichi:2007tf}.} has been proposed to obtain a spherically symmetric LQG-BH \cite{Alonso-Bardaji:2021yls, Alonso-Bardaji:2022ear}. For the sake of convenience, we will refer this novel polymerisation scheme as the Alonso-Bardaji-Brizuela-Vera (ABBV) scheme, and thus this novel LQG-BH model as the ABBV BH. Like many other LQG-BH models, the quantum gravity effects introduced in this novel ABBV BH model removes the classical singularity. A more significant advancement is that in this model of \cite{Alonso-Bardaji:2021yls, Alonso-Bardaji:2022ear}, the modified constraint algebra exhibits closure. Thus, the system offers a reliable and clear geometric representation that remains consistent, covariant, and unambiguous, regardless of the gauge choice on the phase space\footnote{More recently, the topic of BHs and covariance in effective quantum gravity has been discussed in \cite{Zhang:2024khj}. In this work, two LQG-BH solutions with general covariance were constructed.}. A multitude of research have investigated various aspects of this model. Solar system test constraints are discussed in \cite{Chen:2023bao}. The potential detection of quantum gravity effects using eccentric extreme mass-ratio inspirals (EMRIs) is explored in \cite{Fu:2024cfk}, similar work for a LQG inspired rotating black hole can be found in \cite{Zi:2024jla}. Gravitational lensing and optical characteristics are studied in \cite{Soares:2023uup, Junior:2023xgl, Balali:2023ccr}. 

The QNMs of the ABBV BH have already been devoted to exploring the possibility of probing the quantum gravity effects through GWs \cite{Fu:2023drp, Moreira:2023cxy, Bolokhov:2023bwm,Gingrich:2024tuf}. The work \cite{Fu:2023drp} pioneered the analysis of QNM spectra for scalar and electromagnetic perturbations in this geometry using the pseudo-spectral method (PSM). Their work revealed two universal features shared by quantum-corrected and modified gravity models: (i) overtone outbursts and (ii) oscillatory mode behavior. Notably, scalar perturbations exhibited faster decay rates with higher oscillation counts compared to electromagnetic counterparts, while maintaining the Schwarzschild power-law tail structure unaffected by LQG corrections in time-domain profiles. Subsequent work by Moreira et al. \cite{Moreira:2023cxy} independently confirmed these results using Leaver's method with Nollert improvement, extending calculations to higher overtones. The author in \cite{Bolokhov:2023bwm} further generalized these studies by including massless/massive scalar, electromagnetic and Dirac fields. Their analysis demonstrated that the overtone behavior is governed by near-horizon geometry, while the fundamental mode is localized at the potential barrier peak. Specially, massive fields support arbitrarily long-lived modes absent in massless cases. Additionally, they derived an analytical eikonal formula for QNMs and its post-eikonal extension as a $1/l$ expansion, where $l$ is the multipole moment. The study presented in \cite{Gingrich:2024tuf} investigated gravitational perturbations through an effective framework where quantum corrections are modeled via Einstein equations coupled to an anisotropic perfect fluid. The research identified universal characteristics, such as overtone outbursts and oscillatory behavior, within these gravitational perturbations. Additionally, the authors investigated the extent of isospectrality violation in the quasi-normal mode (QNM) spectra of GWs. 

Building upon their prior methodology developed in Refs. \cite{Alonso-Bardaji:2021yls, Alonso-Bardaji:2022ear}, Alonso-Bardaji et al. \cite{Alonso-Bardaji:2023niu} proposed a significant extension of the ABBV model by incorporating electric charge and a cosmological constant. This enhanced framework enables the model to describe the dynamic spacetimes and the asymptotic boundaries. Furthermore, this extension aligns seamlessly with the generalized laws of BH thermodynamics, particularly the interplay between entropy, charge, and cosmological constant in the fourth law. In a complementary study, Borges et al. \cite{Borges:2023fub} established two critical results: (i) The Cauchy horizon is rigorously shown to reside within the transition surface, suggesting a quantum-gravitational mechanism for stabilizing causal structures and potentially resolving the classical instability associated with Cauchy horizons; (ii) The existence of limiting quantum gravitational states with vanishing surface gravity is demonstrated, which may correspond to the final stages of black hole evaporation or the emergence of remnant states at the Planck scale.

Our study focuses on analyzing the QNMs of scalar and Dirac fields in the spacetime of charged ABBV BH\footnote{Gravitational and electromagnetic perturbations in this charged ABBV framework present significant technical challenges, and we defer their investigation to future work.}. Especially, we work out the overtone modes of the Dirac field using PSM, which were absent in the previous work of netural ABBV BH \cite{Bolokhov:2023bwm}. In our present work, we pioneer the study of the combined influence of charge and effective quantum gravitational effects on QNMs. A noteworthy finding is that while individual parameters can induce overtone outbursts, the combined effect reveals that charge suppresses such outbursts, necessitating observation at higher overtones. The structure of our paper is as follows. In Section \ref{sec.scl feld over LQG}, we provide a brief review of the charged ABBV BH and the dynamics of the scalar field and Dirac field. Section \ref{sec:QNMs} presents a systematic study of the properties of QNMs. Finally, conclusions and further discussions are given in Section \ref{sec:conclusion}.

\section{Scalar and Dirac field over the charged ABBV black hole}
\label{sec.scl feld over LQG}
In this section, we firstly present a brief overview of the charged ABBV BH, originally proposed in Ref.~\cite{Alonso-Bardaji:2023niu}. The charged ABBV BH represents an extension of the neutral ABBV BH model detailed in Refs.~\cite{Alonso-Bardaji:2021yls, Alonso-Bardaji:2022ear}. Subsequently, we derive the equations of motion (EOM) governing the dynamics of scalar and Dirac fields within this spacetime. Then we analyze the properties of the resulting effective potentials, highlighting their physical implications.

Following the framework established in Refs. \cite{Alonso-Bardaji:2021yls, Alonso-Bardaji:2022ear}, the authors of Ref.~\cite{Alonso-Bardaji:2023niu} proposed an effective LQG-corrected Hamiltonian incorporating electric charge as follows \footnote{In this analysis, we restrict our consideration to asymptotically flat spacetimes and thus set the cosmological constant to zero, as in \cite{Borges:2023fub}.}:
\begin{eqnarray}\label{hamiltonian}
H_{\rm eff}
&= -\dfrac{1}{2G\gamma\sqrt{1+\gamma^2\delta_b^2}}\left[  \left(\dfrac{\sin(\delta_b b)}{\delta_b} +\dfrac{\gamma^2 \delta_b}{\sin{(\delta_b b)}}- \dfrac{\gamma^2 \delta_b Q^2}{\sin( \delta_b b)p_c }\right)p_b + 2cp_c\cos(\delta_b b)\right]\,.
\end{eqnarray}
Here, $b$, $p_b$, $c$ and $p_c$ are the conjugate variables satisfying the Poisson brackets as $\{b,p_b\} = G\gamma$ and $\{c,p_c\} = 2G\gamma$. Additionally, $Q$ is a constant associated with the charge of the BH, $\gamma$ is the Immirzi parameter, and $\delta_b$ is the polimerization parameter. The classical limit is obtained by taking $\delta_b \rightarrow 0$. It is important to emphasize that as $Q\rightarrow 0$, the effective Hamiltonian presented above reduces to the neutral case \cite{Alonso-Bardaji:2021yls, Alonso-Bardaji:2022ear}.

From the effective LQG-corrected Hamiltonian presented in Eq.~\eqref{hamiltonian}, the dynamical equations can be derived. The solution to these equations results in the static metric below\footnote{For a detailed derivation, see Ref.~\cite{Alonso-Bardaji:2023niu}, where the results were first derived, and Ref.~\cite{Borges:2023fub} for subsequent developments.}:
\begin{equation}
d{s^2} =  - f(r)d{\tau ^2} +\frac{1}{\left( {1 - \frac{{{r_0}}}{m}g\left( r \right)} \right)f(r)}d{r^2} + {r^2}d{\Omega ^2}\,.
\label{ds}
\end{equation}
The functions $f(r)$ and $g(r)$ in the above equation are defined as follows, respectively:
\begin{eqnarray}
&&
f(r)=1 - \frac{{2m}}{r} + \frac{{{Q^2}}}{{{r^2}}}\,,
\label{fr}
\
\\
&&
g\left( r \right) = \frac{{\frac{{2m}}{r} - \frac{{{Q^2}}}{{{r^2}}}}}{{1 + \sqrt {1 - \frac{{b_0^2{Q^2}}}{{\left( {b_0^2 - 1} \right){m^2}}}} }}\,,
\label{gr}
\end{eqnarray}
where $b_0$ is defined as ${b_0}\equiv \sqrt {1 + {\gamma ^2}\delta _b^2} $. The constant of motion $m$ is associated with the Komar mass at spatial infinity \cite{Borges:2023fub,Alonso-Bardaji:2022ear}. This BH geometry is asymptotically flat at infinity and its horizons are located at
\begin{equation}
r_h^ \pm  = m\left( {1 \pm \sqrt {1 - \frac{{{Q^2}}}{{{m^2}}}} } \right)\,.
\label{rh}
\end{equation}
It is noted that the horizons of this charged ABBV BH are the same as those of the Reissner-Nordstr\"om (RN) BH.

With the introduction of LQG gravity effects, the spacetime singularity is replaced by a transition surface, also known as the bounce, which connects a BH to a white hole (WH). The bounce radius $r_0$ is given by \cite{Alonso-Bardaji:2023niu,Borges:2023fub}
\begin{equation}
	{r_0} = \frac{{\left( {b_0^2 - 1} \right)}}{{b_0^2}}m\left( {1 + \sqrt {1 - \frac{{b_0^2{Q^2}}}{{\left( {b_0^2 - 1} \right){m^2}}}} } \right)\,,
\end{equation}
with 
\begin{equation}
\left| Q \right| \le m\frac{{\sqrt {b_0^2 - 1} }}{{{b_0}}}\,.
\label{eqn:Q upper limit}
\end{equation}
It is found that the Cauchy horizon is always hidden within the bounce radius ${r_0}$, satisfying the inequality $r_h^ -  < {r_0} < r_h^ + $ \cite{Borges:2023fub}. 

Next, we investigate the response of the charged ABBV black hole to external perturbations generated by a scalar field $\Psi$ and a Dirac field $\Upsilon$. The dynamics of these fields are governed by the following equations:
\begin{eqnarray} 
&&\frac{1}{{\sqrt { - g} }}{\partial _\nu }\left( {{g^{\mu \nu }}\sqrt { - g} {\partial _\mu }\Psi } \right) -\mu ^2 \Psi= 0\,, 
\label{KG-eq} 
\
\\
&& {\gamma ^\alpha }\left( {\frac{\partial }{{\partial {{x}^\alpha }}} - {\Gamma _\alpha }} \right)\Upsilon  = 0\,,
    \label{Dirac-eq}
\end{eqnarray}
where $g_{\mu \nu}$ is the background metric of the charged ABBV spacetime, $\mu$ denotes the scalar field mass, ${\gamma ^\alpha }$ are the gamma matrices, and ${\Gamma _\alpha }$ represents the spin connection in the tetrad formalism \cite{Brill:1957fx}.

Through a separation of variables in the static spherically symmetric background, both equations can be reduced to a unified Schr\"odinger-like wave equation:
 \begin{equation}
 \frac{{{\partial ^2}\psi }}{{\partial r_*^2}} + \left( {{\omega ^2} - V_i} \right)\psi  = 0\,,
 \label{sch-eq}
\end{equation}
where $r_*$ is the tortoise coordinate defined by
\begin{equation}
\frac{{d{r_*}}}{{dr}} = {\left( {1 - \frac{{2m}}{r} + \frac{{{Q^2}}}{{{r^2}}}} \right)^{ - 1}}\sqrt {{{\left( {1 - \frac{{{r_0}}}{m}g\left( r \right)} \right)}^{ - 1}}}\,.
\label{drast-dr}
\end{equation} 
The effective potentials $V_i$ exhibit distinct structures for scalar and Dirac fields. For the scalar perturbation:
\begin{equation}
    V_{\text{scalar}} = f(r)\left(\frac{{l\left( {l + 1} \right)}}{{{r^2}}}+\mu^2\right) + \frac{{f(r)\left[ {  2\left( {m - {r_0}g(r)} \right)f'(r) - {r_0}f(r)g'(r)} \right]}}{{2mr}}\,,
    \label{V effect potential}
\end{equation}
while for the Dirac field, the spin-dependent potentials are:
\begin{equation}
    V_{\text{Dirac}}^ \pm  = f\left( r \right)\frac{{{k^2}}}{{{r^2}}} \pm \frac{{k\sqrt {f\left( r \right)\left( {1 - \frac{{{r_0}}}{{g\left( r \right)}}} \right)} \left( { - 2f\left( r \right) + rf'\left( r \right)} \right)}}{{2{r^2}}}\,.
    \label{V_Dirac field}
\end{equation}
Here, the multipole quantum numbers $l=0,1,2...$ for the scalar field, and $k=1,2,3...$ for the Dirac field.  $V_{\text{Dirac}}^+$ and $V_{\text{Dirac}}^-$ correspond to spin-up and spin-down fermionic modes, respectively. These potentials are interrelated via a Darboux transformation \cite{Konoplya:2011qq,Zinhailo:2019rwd}, which allows them to be mutually transformed. For brevity, this work focuses exclusively on the $V_{\text{Dirac}}^ +$ case.

\begin{figure}[H]
    \centering
    \begin{minipage}{0.45\textwidth}
        \centering
        \includegraphics[width=\linewidth]{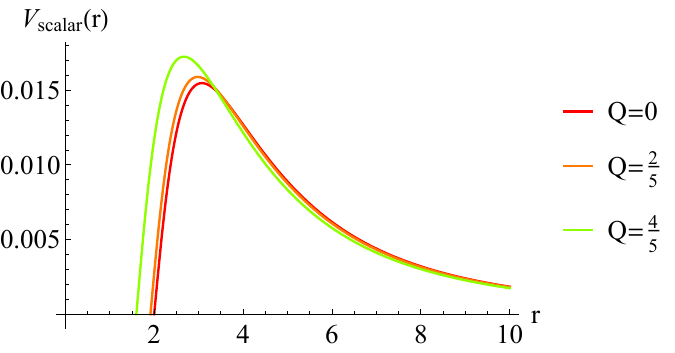}
    \end{minipage}
    \hfill
    \begin{minipage}{0.45\textwidth}
        \centering
        \includegraphics[width=\linewidth]{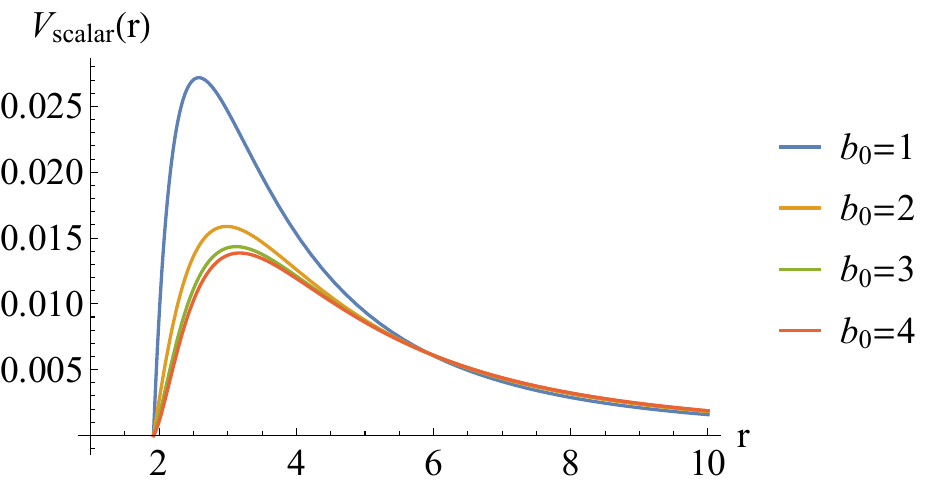}
    \end{minipage}
    \caption{The effective potential of massless scalar field as a function $r$ with $l=0$, while considering various values for both the quantum parameter $b_0$ and the charge $Q$. In the left panel, we fix the quantum parameter $b_0=2$, and then take the charge as $Q = 0,\frac{2}{5},\frac{4}{5}$. Conversely, in the right panel, we set $Q=\frac{2}{5}$, and then the quantum parameter as $b_0=1, 2, 3, 4$.}
    \label{fig:potential for l=0}
\end{figure}

 \begin{figure}[H]
    \centering
    \begin{minipage}{0.45\textwidth}
        \centering
        \includegraphics[width=\linewidth]{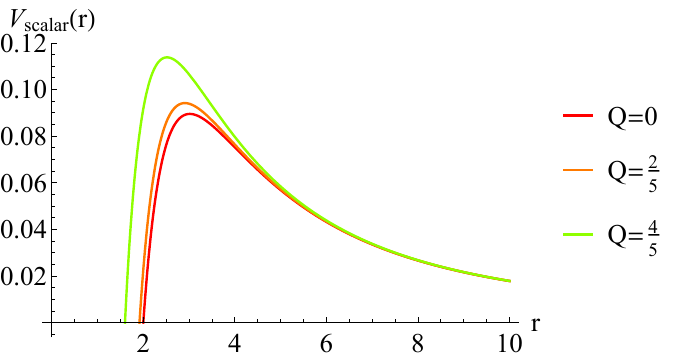}
    \end{minipage}
    \hfill
    \begin{minipage}{0.45\textwidth}
        \centering
        \includegraphics[width=\linewidth]{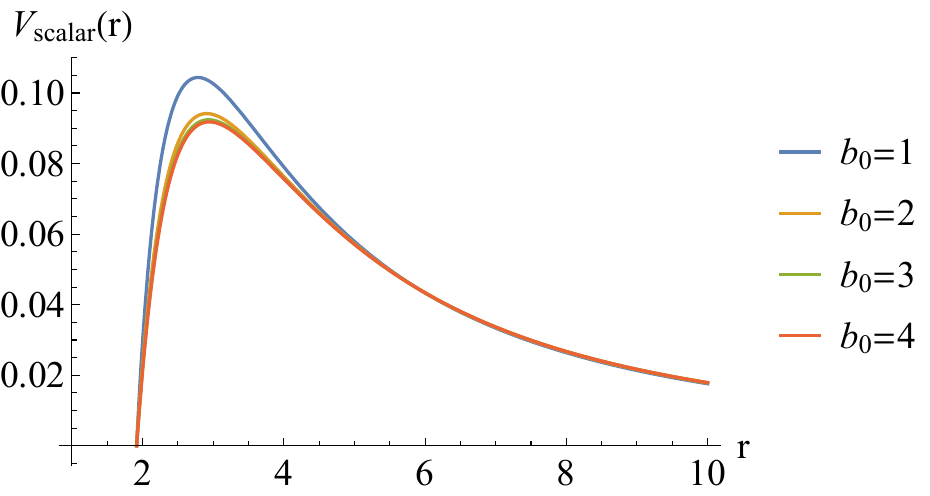}
    \end{minipage}
    \caption{The effective potential of massless scalar field as a function $r$ with $l=1$, while considering various values for both the quantum parameter $b_0$ and the charge $Q$. In the left panel, we fix the quantum parameter $b_0=2$, and then take the charge as $Q = 0, \frac{2}{5}, \frac{4}{5}$. Conversely, in the right panel, we set $Q=\frac{2}{5}$, and then take the quantum parameter as $b_0=1, 2, 3, 4$.}
    \label{fig:potential for l=1}
\end{figure}

We begin by analyzing the effective potential $V_{\text{scalar}}$ for a massless scalar field, which is structurally decomposed into two distinct components as defined in Eq.\eqref{V effect potential}. The first term is the centrifugal potential, associated with the multipole quantum number $l$. The centrifugal potential prevents the wave from approaching the center, forming a barrier in regions farther away from it. The second term is the gravitational potential, related to the gravitational field or the geometry of spacetime, which typically introduces additional attraction or barriers, influencing the wave's decay characteristics. Notably, the quantum parameter $b_0$ appears only in the second term, making the effects of quantum gravity most significant when $l=0$. As $l$ increases, the influence of the multipole quantum number tends to overshadow the quantum gravity effects---a phenomenon observed in our previous work \cite{Song:2024kkx,Zhang:2024nny,Gong:2023ghh,Fu:2023drp}. Therefore, to better explore the effects of quantum gravity and charge $Q$, we focus primarily on the $l=0$ case and then briefly discuss the $l=1$ case.

Fig.\ref{fig:potential for l=0} and Fig.\ref{fig:potential for l=1} illustrate the effective potential $V_{\text{scalar}}$ as a function of $r$ for varying $Q$ and $b_0$ in the case of $l=0$ and $l=1$, respectively\footnote{Unless otherwise specified, we will set $m=1$ throughout this paper.}. The effective potential remains positive throughout, indicating the stability of the charged ABBV BH under scalar perturbations. Moreover, it is observed that the charge parameter $Q$ and the quantum parameter $b_0$ predominantly affect the behaviors of the effective potential near the horizon. It is a significant factor in inducing the so-called overtone’s outburst, as will be demonstrated below.

Before proceeding, we provide a qualitative analysis of the properties of the effective potential $V_{\text{scalar}}$, focusing on its implications for the QNMs characteristics discussed in subsequent sections. 
We first examine the case with fixed $b_0$ (see the left plots in Fig.\ref{fig:potential for l=0} and Fig.\ref{fig:potential for l=1}). From these plots, it is evident that the relative change for $Q=\frac{2}{5}$ or $Q=\frac{4}{5}$ (relative to $Q=0$) is larger for $l=1$ than for $l=0$. This arises from the inclusion of the charge parameter in the centrifugal potential, where the increase in $l$ results in a combined effect of $l$ and $Q$, which induces such significant deviations. Next, we consider the case with the charge parameter $Q$ fixed (see the right plots in Fig.\ref{fig:potential for l=0} and Fig.\ref{fig:potential for l=1}). Compared to the case of $b_0=1$, the relative change with $b_0>1$ is more pronounced for $l=0$ than for $l=1$. This is because, as $l$ increases, the influence of the multipole quantum number tends to overshadow the effects of quantum gravity.
\begin{figure}[H]
    \centering
    \begin{minipage}{0.45\textwidth}
        \centering
        \includegraphics[width=\linewidth]{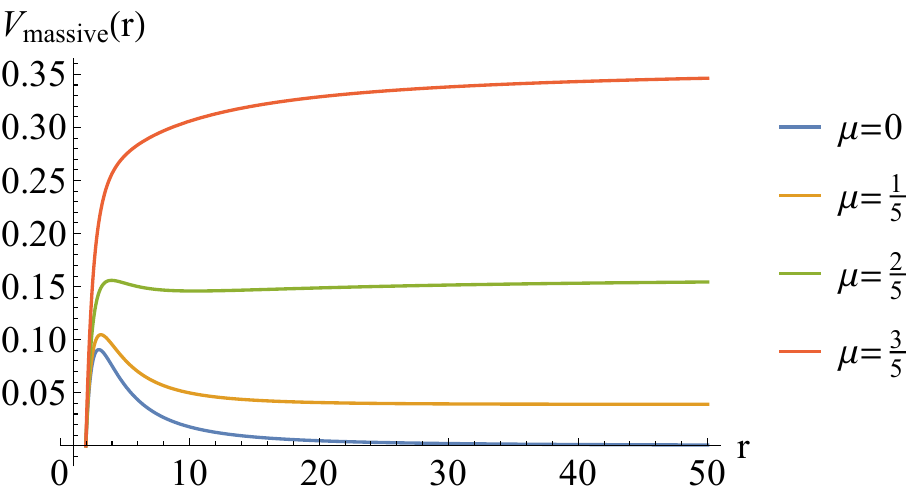}
    \end{minipage}
    \hfill
    \begin{minipage}{0.45\textwidth}
        \centering
        \includegraphics[width=\linewidth]{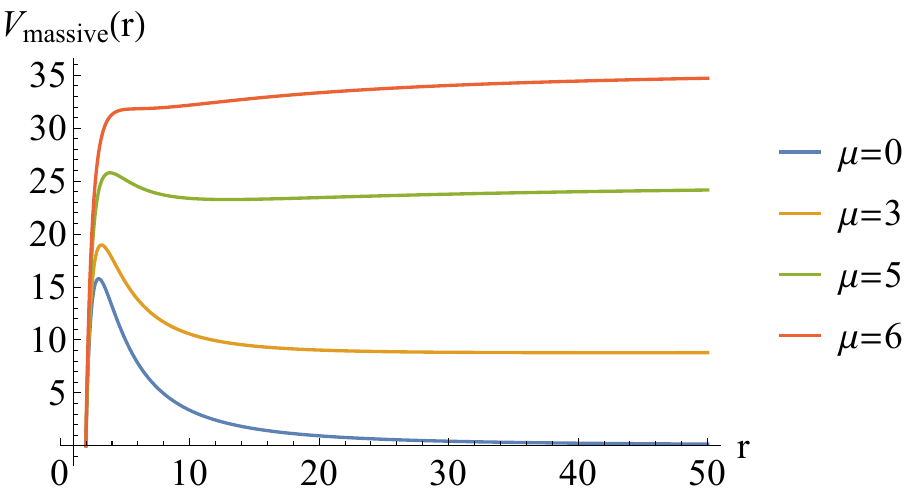}
    \end{minipage}
    \caption{The effective potential of massive scalar field as a function $r$ with $Q=\frac{1}{5}, b_0=2$, while considering various values for mass of saclar fields $\mu$. In the left panel, we set the $l=1$, and then take the mass of saclar fields as $\mu = 0,\frac{1}{5},\frac{2}{5},\frac{3}{5}$. In the right panel, we set $l=20$, and then the quantum parameter as $\mu=0, 3, 5, 6$.}
    \label{fig:massive scalar potential for l=1,20}
\end{figure}

We now analyze the properties of the effective potential for the massive scalar field. As evident from the potential $V_{\text{scalar}}$ (Eq.\eqref{V effect potential}), the asymptotic behavior of $V_{\text{scalar}}$ differs significantly from the massless case: at spatial infinity ($r\rightarrow\infty$), the potential approaches a constant value $\mu^2$, in contrast to the vanishing behavior ($V_{\text{scalar}}\rightarrow 0$) for massless fields. This distinction is also explicitly demonstrated in Fig. \ref{fig:massive scalar potential for l=1,20}. The inclusion of the mass term $\mu^2$ raises the overall height of the potential, particularly in far-field regime ($r\gg 2M$), where it may form a more pronounced barrier. This affects the reflection and transmission properties of the waves, thereby altering both the oscillation frequencies and damping times of the QNMs.

\begin{figure}[H]
    \centering
    \begin{minipage}{0.45\textwidth}
        \centering
        \includegraphics[width=\linewidth]{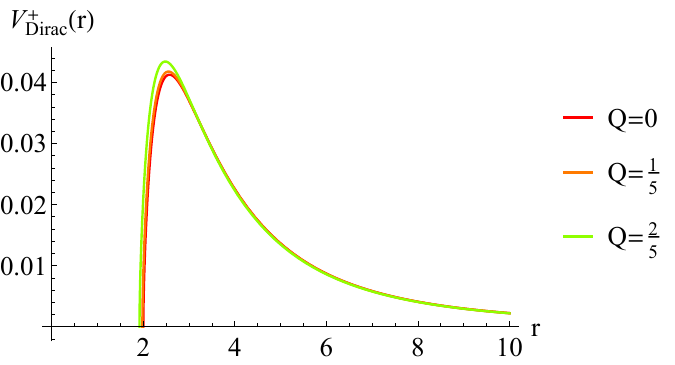}
    \end{minipage}
    \hfill
    \begin{minipage}{0.45\textwidth}
        \centering
        \includegraphics[width=\linewidth]{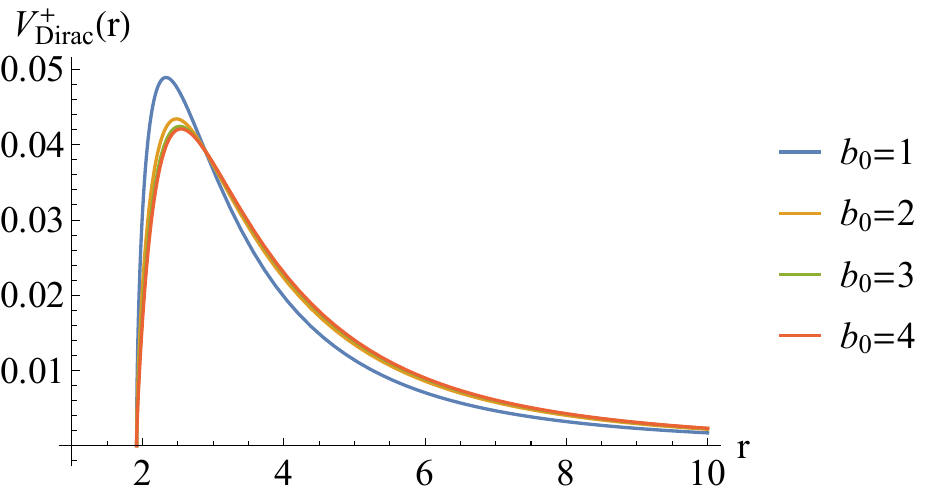}
    \end{minipage}
    \caption{The effective potential of Dirac fiels as a function $r$ with $k=1$, while considering various values for both the quantum parameter $b_0$ and the charge $Q$. In the left panel, we fix the quantum parameter $b_0=2$, and then take the charge as $Q=0, \frac{1}{5}, \frac{2}{5}$. Conversely, in the right panel, we set $Q=\frac{2}{5}$, and then take the quantum parameter as $b_0=1, 2, 3, 4$.}
    \label{fig:Dirac potential for l=1}
\end{figure}

For the Dirac field, both the centrifugal and gravitational potential components exhibit explicit dependence on the multipole quantum number $k$. As illustrated in Fig.\ref{fig:Dirac potential for l=1}, the effective potential $V_{\text{Dirac}}^+$ displays a universally positive profile, indicating the stability of the charged ABBV black hole under Dirac field perturbations. While parametric tuning of the LQG-corrected parameter $b_0$ or the charge $Q$ induces qualitative similarities in the potential change between Dirac and scalar fields, the amplitude of variation in $V_{\text{Dirac}}^+$ is systematically suppressed compared to its scalar counterpart.

\section{Quasinormal modes}
\label{sec:QNMs}
In this section, we investigate the properties of QNMs generated by scalar and Dirac field perturbations in the context of charged ABBV BH. The QNMs are obtained by solving an eigenvalue problem under specific boundary conditions: the wave function must represent a purely outgoing wave at spatial infinity ($ r_*\rightarrow+\infty $) and a purely ingoing wave at the event horizon ($ r_*\rightarrow-\infty $). Mathematically, this is expressed as:
\begin{equation}
    \psi  \sim {e^{ \pm i\omega {r_*}}},  \,\,\,\,\,\,\,\,\,\,\,\,  {r_*} \to  \pm \infty\,.
    \label{boundary conditions}
\end{equation}
These boundary conditions reflect the BH's response to a transient perturbation, capturing the dynamics after the perturbing source has dissipated \cite{Konoplya:2011qq,Berti:2009kk,Kokkotas:1999bd}.

A variety of methods have been developed to calculate QNMs, including the WKB method \cite{Ferrari:1984zz,Schutz:1985km,Iyer:1986np,Iyer:1986nq,Konoplya:2003ii,Matyjasek:2017psv}, the asymptotic iteration method (AIM) \cite{Ciftci:2005xn,Cho:2009cj,Cho:2011sf}, the Horowitz-Hubeny method \cite{Horowitz:1999jd}, the continued fraction method (CFM) \cite{Leaver:1985ax}, and the PSM \cite{boyd2001chebyshev,Jansen:2017oag}. In this study, we primarily utilize the PSM to calculate the QNM spectra, given its effectiveness as a robust numerical tool \cite{boyd2001chebyshev,Jansen:2017oag,Wu:2018vlj,Fu:2018yqx,Fu:2022cul,Gong:2023ghh,Fu:2023drp,Zhang:2024nny,Liu:2021fzr,Destounis:2021lum,Jaramillo:2021tmt,Xiong:2021cth,Song:2024kkx}, particularly in resolving overtone modes \cite{Fu:2023drp,Gong:2023ghh,Zhang:2024nny}. To rigorously validate the reliability of our PSM-based QNM analysis, we compare the PSM results with those obtained from the higher-order WKB method for the fundamental modes. A comprehensive discussion of this comparison is presented in Appendix \ref{App-WKB}.

To determine the QNM spectra using the PSM, two key steps are essential: first, working in Eddington-Finkelstein coordinates; second, discretizing the differential equations and solve the resulting generalized eigenvalue problem. The choice of Eddington-Finkelstein coordinates simplifies the boundary condition setup. To this end, we introduce the following transformations \footnote{In the case of Dirac field, we use the transformations: $r = {r_h}/{1-u^2}$, taken from \cite{Stashko:2024wuq}.}:
\begin{equation}
r = \frac{{{r_h}}}{u}\,,~~~{\rm{and}}\,~~~\psi  \rightarrow {e^{i\omega {r_*}\left( u \right)}}\psi \,.
\label{tran}
\end{equation}
These transformations allow us to impose only the outgoing boundary condition at infinity. Note that, to simplify notation, we will henceforth abbreviate the outer event horizon $r_h^ +$ as $r_h$. With the above considerations, the wave equation \eqref{drast-dr} is transformed into:
\begin{equation}
    {\alpha _0}\psi '' + {\beta _0}\psi ' + {\gamma _0}\psi  = 0\,.
\end{equation}
The specific forms of the coefficients ${\alpha _0}$, ${\beta _0}$, and ${\gamma _0}$ for both the massless scalar field and the Dirac field are detailed in Appendix \ref{Appendix:equation for PSM}.

Next, we discretize the continuous variables using a set of Chebyshev grids and express the functions in terms of Lagrange cardinal functions. The Chebyshev grid points and the Lagrange cardinal functions are defined as follows:
\begin{eqnarray}
&&
{x_i} = \cos \left( {\frac{i}{N}\pi } \right)\,,
\
\\
&&
{C_j}(x) = \prod\limits_{i = 0,i \ne j}^N {\frac{{x - {x_i}}}{{{x_j} - {x_i}}}} ,\, \, \, \, \, i = 0,...,N\,.
\end{eqnarray}
Following these operations, the wave equation is reduced to a generalized eigenvalue problem of the form:
\begin{equation}
    \left( {{M_0} + \omega {M_1}} \right)\psi  = 0\,,
\end{equation}
where $M_i(i=0,1)$ represents a linear combination of the derivative matrices. The QNFs are then determined by solving this generalized eigenvalue equation.

Next, we will explore the properties of the scalar field and the Dirac field over the charged ABBV BH, respectively.

\subsection{Scalar field case}\label{subsec-Scalar}

\begin{figure}[H]
    \centering
    \begin{minipage}{0.48\textwidth}
        \centering
        \includegraphics[width=\linewidth]{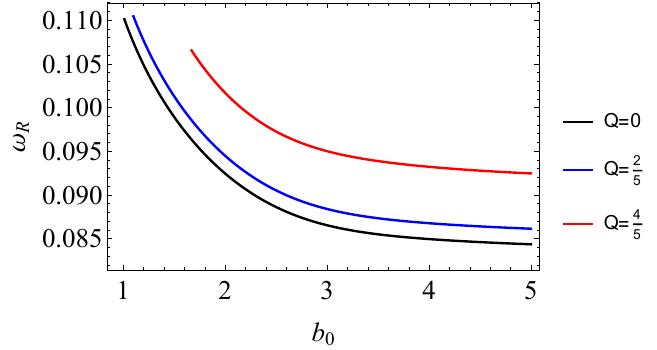}
    \end{minipage}
    \hfill
    \begin{minipage}{0.48\textwidth}
        \centering
        \includegraphics[width=\linewidth]{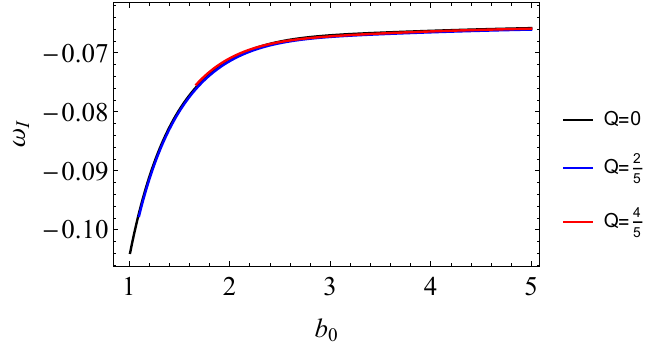}
    \end{minipage}
    \caption{For massless scalar field, QNFs of fundamental modes are presented as a function of the quantum parameter $b_0$ for $l=0$, while considering various vales of the charge, specifically $Q=0, \frac{2}{5}, \frac{4}{5}$.}
    \label{fig:FFvsb0}
\end{figure}

We begin by studying the massless scalar field case with $l=0$, where only the gravitational potential remains. Fig.\ref{fig:FFvsb0} illustrates the real part, $\omega_R$, and the imaginary part, $\omega_I$, of the fundamental modes as functions of the quantum parameter $b_0$ for various values of the charge parameter $Q$. When $Q$ is fixed, an increase in $b_0$ leads to a monotonic decrease in $\omega_R$ and a monotonic increase in  $\omega_I$. This behavior is consistent with the uncharged case (the black line for $Q=0$ and also see Refs.\cite{Fu:2023drp,Bolokhov:2023bwm}). These observations suggest that the LQG effect reduces the oscillation behavior but simultaneously causes a slower decay of the modes, independent of the charge parameter $Q$. 

\begin{figure}[H]
    \centering
    \begin{minipage}{0.45\textwidth}
        \centering
        \includegraphics[width=\linewidth]{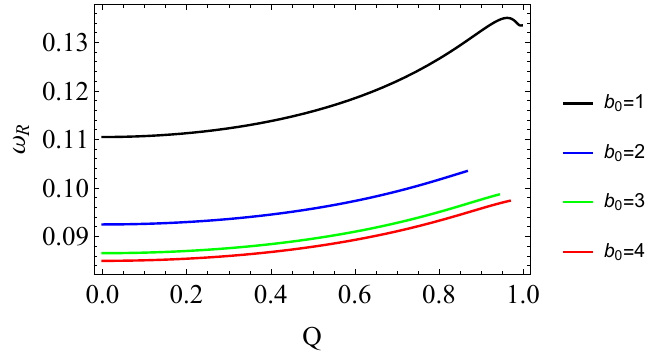}
    \end{minipage}
    \hfill
    \begin{minipage}{0.45\textwidth}
        \centering
        \includegraphics[width=\linewidth]{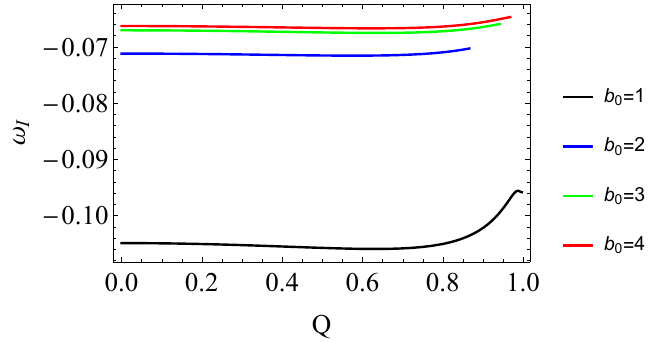}
    \end{minipage}
    \caption{For massless scalar field, QNFs of fundamental modes are presented as a function of the charge $Q$ for $l=0$, while considering various values of the quantum parameter, specifically $b_0=1,2,3,4$.}
    \label{fig:FFvsQ}
\end{figure}

In Fig.\ref{fig:FFvsQ}, we fix the quantum parameter $b_0$ and investigate how the fundamental modes change with $Q$. When $b_0\neq 1$, the real part of the frequency, $\omega_R$, shows a monotonic increase as $Q$ increases (left plot in Fig.\ref{fig:FFvsQ}), whereas the imaginary part, $\omega_I$, remains nearly constant across a broad range of $Q$ (see the right plots in Fig.\ref{fig:FFvsQ}). However, for large values of $Q$, $\omega_I$ shows a slight increase. These findings suggest that the charge $Q$ in this model enhances the oscillatory behavior while having minimal effect on the decay of the modes. 

Different from the previous effective quantum gravity models \cite{Song:2024kkx,Fu:2022cul,Fu:2023drp,Gong:2023ghh,Zhang:2024nny}, we do not observe non-monotonic behavior of the QNFs with respect to either the quantum parameter $b_0$ or the charge parameter $Q$. However, it is worth noting that when $b_0=1$ corresponding to the RN scenario, both $\omega_R$ and $\omega_I$ as functions of $Q$ display non-monotonic behavior as the extremal BH is approached, i.e., as $Q$ tends to $1$ (see the black line in Fig.\ref{fig:FFvsQ}). Therefore, concerning non-monotonic behaviors, $b_0$ and $Q$ play mutually inhibiting roles.

\begin{figure}[H]
    \centering
    \begin{minipage}{0.48\textwidth}
        \centering
        \includegraphics[width=\linewidth]{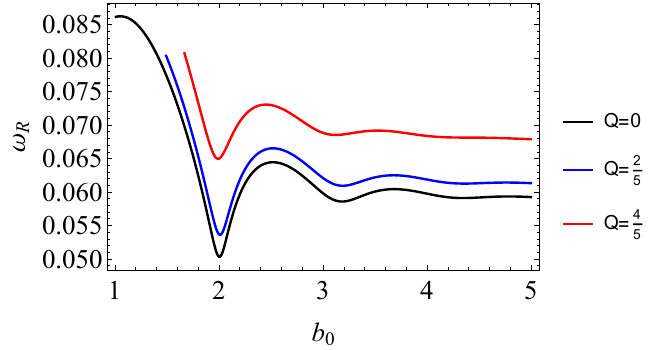}
    \end{minipage}
    \hfill
    \begin{minipage}{0.48\textwidth}
        \centering
        \includegraphics[width=\linewidth]{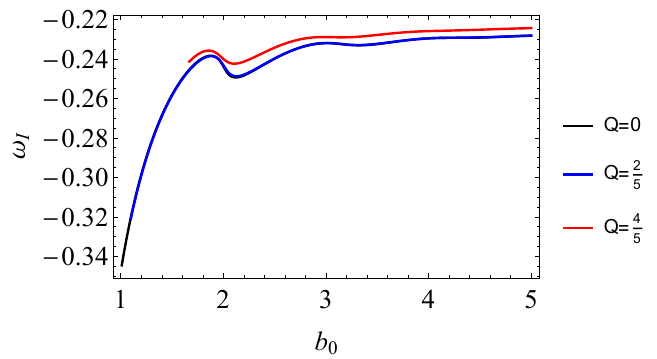}
    \end{minipage}
    \caption{For massless scalar field, QNFs of the first overtone ($n=1$) with $l=0$ are presented as a function of the quantum parameter $b_0$ for the values of the charge $Q=0, \frac{2}{5}, \frac{4}{5}$. The left and right panels show the real and imaginary parts, respectively.}
    \label{fig:omegaRvsb0n1}
\end{figure}
\begin{figure}[H]
    \centering
    \begin{minipage}{0.48\textwidth}
        \centering
        \includegraphics[width=\linewidth]{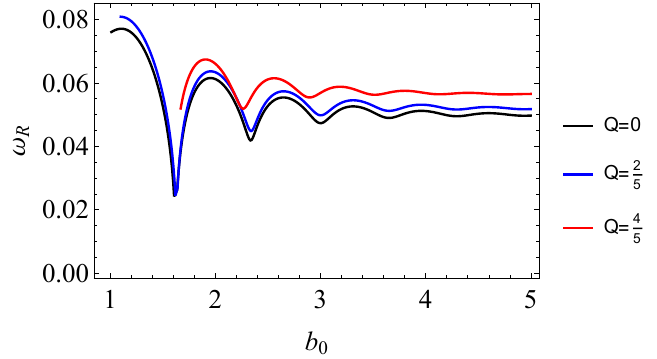}
    \end{minipage}
    \hfill
    \begin{minipage}{0.48\textwidth}
        \centering
        \includegraphics[width=\linewidth]{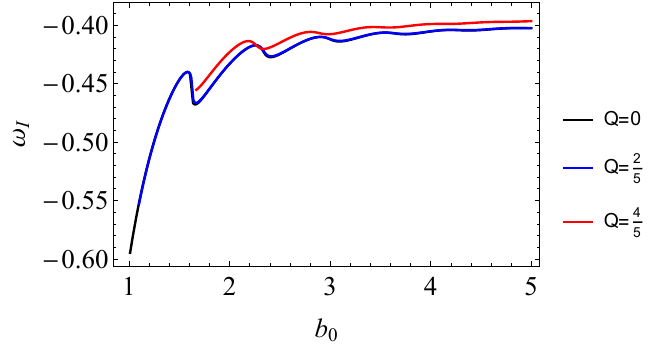}
    \end{minipage}
    \caption{For massless scalar field, QNFs of the second overtone ($n=2$) with $l=0$ are presented as a function of the quantum parameter $b_0$ for the values of the charge $Q=0, \frac{2}{5}, \frac{4}{5}$. The left and right panels show the real and imaginary parts, respectively.}
    \label{fig:omegaRvsb0n2}
\end{figure}

We proceed to study the properties of the overtones with $l=0$. Fig.\ref{fig:omegaRvsb0n1} and Fig.\ref{fig:omegaRvsb0n2} show the QNFs as functions of the quantum parameter $b_0$ for various values of $Q$, corresponding to the first two overtones (Fig.\ref{fig:omegaRvsb0n1} for $n=1$ and Fig.\ref{fig:omegaRvsb0n2} for $n=2$, respectively). It is evident that the quantum gravity effects trigger the outburst of overtones, resulting in noticeable changes in the QNFs compared to those of Schwarzschild or RN BHs. This phenomenon has been widely observed in the modified gravity theory and effective quantum gravity models \cite{Berti:2005ys,Konoplya:2022pbc,Berti:2018vdi,Fu:2022cul,Fu:2023drp,Gong:2023ghh,Moura:2021eln,Moura:2021nuh,Moura:2022gqm,Lin:2024ubg,Ghosh:2022gka,Konoplya:2023aph,Zinhailo:2023xdz,Zhang:2024nny,Song:2024kkx,Konoplya:2024lch,Stashko:2024wuq,Dubinsky:2024nzo,Zinhailo:2024kbq}. Following the initial outburst of overtones, we observe an oscillatory behavior. This oscillation gradually becomes weak as the quantum parameter $b_0$ increases. This pattern has been observed in RN-BH \cite{Berti:2003zu,Jing:2008an} and other effective quantum gravity corrected BH \cite{Fu:2023drp,Gong:2023ghh,Zhang:2024nny} and is likely associate with the extremal effect.

\begin{figure}[H]
    \centering
    \begin{minipage}{0.48\textwidth}
        \centering
        \includegraphics[width=\linewidth]{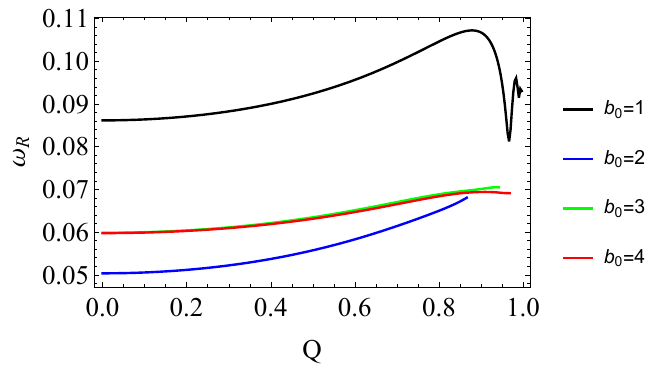}
    \end{minipage}
    \hfill
    \begin{minipage}{0.48\textwidth}
        \centering
        \includegraphics[width=\linewidth]{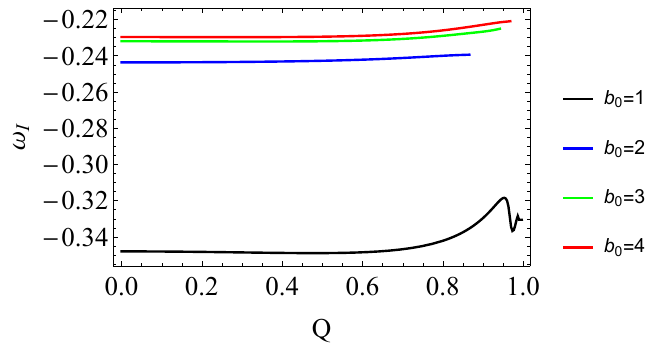}
    \end{minipage}
    \caption{For massless scalar field, QNFs of the first overtone ($n=1$) with $l=0$ are presented as a function of the charge $Q$ for different quantum parameters $b_0=1,2,3,4$. The left and right panels show the real and imaginary parts, respectively.}
    \label{fig:omegaRvsQn1}
\end{figure}

As expected, with increasing $n$, the overtone outburst becomes more pronounced, with smaller quantum parameters capable of triggering this outburst. For fixed $n$, the oscillatory behavior remains consistent across different charge parameters $Q$. Additionally, as $n$ increases, the oscillations become stronger. On the other hand, as $Q$ increases, both the strength of the overtone outburst and the oscillation become weak, suggesting that the charge does not enhance either the overtone outburst or the oscillatory behavior.

\begin{figure}[H]
    \centering
    \begin{minipage}{0.48\textwidth}
        \centering
        \includegraphics[width=\linewidth]{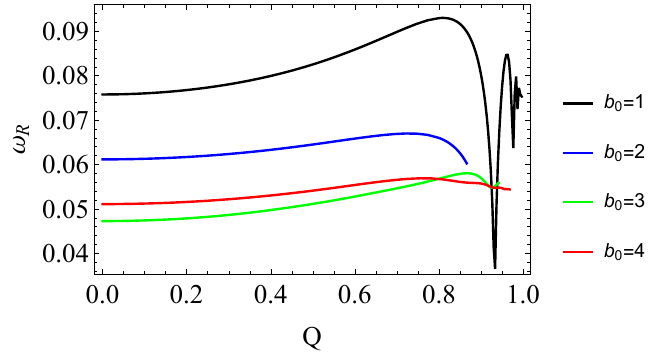}
    \end{minipage}
    \hfill
    \begin{minipage}{0.48\textwidth}
        \centering
        \includegraphics[width=\linewidth]{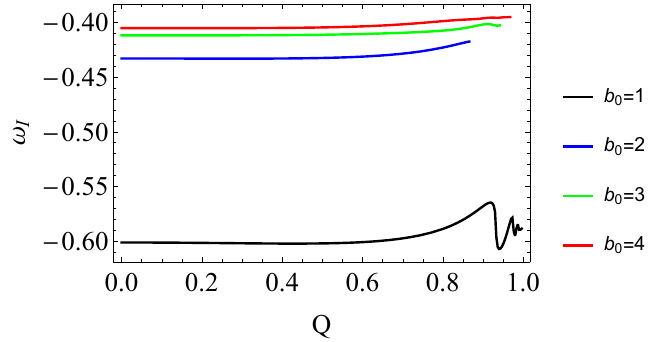}
    \end{minipage}
    \caption{For massless scalar field, QNFs of the second overtone ($n=2$) with $l=0$ are presented as a function of the charge $Q$ for different quantum parameters $b_0=1,2,3,4$. The left and right panels show the real and imaginary parts, respectively.}
    \label{fig:omegaRvsQn2}
\end{figure}

We now study the behavior of the QNFs as functions of $Q$ with a fixed $b_0$ (see Fig.\ref{fig:omegaRvsQn1} and Fig.\ref{fig:omegaRvsQn2}). In the case where $b_0=1$, corresponding to the RN-BH, an overtone outburst with a distinct oscillatory pattern is observed (Fig.\ref{fig:omegaRvsQn1}). As the overtone number $n$ increases, both the intensity of the overtone outburst and the prominence of the oscillations become more pronounced. Nevertheless, upon activation of the quantum parameter $b_0$ ($b_0>1$), the overtone outburst vanishes for the first overtone within the allowed range of $Q$ (see Fig.\ref{fig:omegaRvsQn1}). For the second overtone, once the quantum parameter $b_0$ is activated, the overtone outburst degenerates into non-monotonic behavior (see Fig.\ref{fig:omegaRvsQn2}). In conclusion, the charge $Q$ and the quantum parameter $b_0$ appear to exert mutually suppressive effects on the overtone outburst. The analogous effect on non-monotonic behaviors has also been noted in the preceding subsection.

\begin{figure}[H]
    \centering
    \begin{minipage}{0.48\textwidth}
        \centering
        \includegraphics[width=\linewidth]{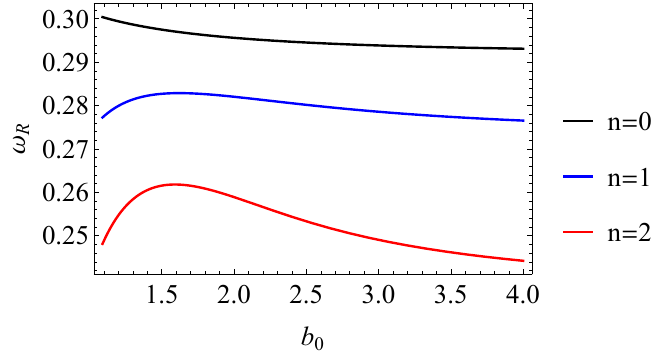}
    \end{minipage}
    \hfill
    \begin{minipage}{0.48\textwidth}
        \centering
        \includegraphics[width=\linewidth]{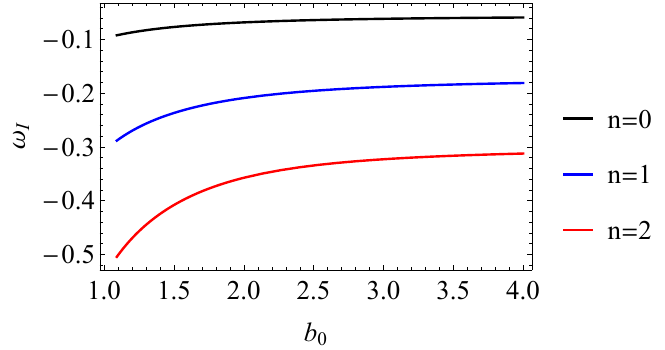}
    \end{minipage}
    \caption{For massless scalar field, QNFs of fundamental mode and the first two overtones with $l=1$ are presented as functions of the quantum parameter $b_0$ for $Q=\frac{2}{5}$. The left and right panels show the real and imaginary parts, respectively.}
    \label{fig:Q2c5 l1n012 Re Im}
\end{figure}

\begin{figure}[H]
    \centering
    \begin{minipage}{0.48\textwidth}
        \centering
        \includegraphics[width=\linewidth]{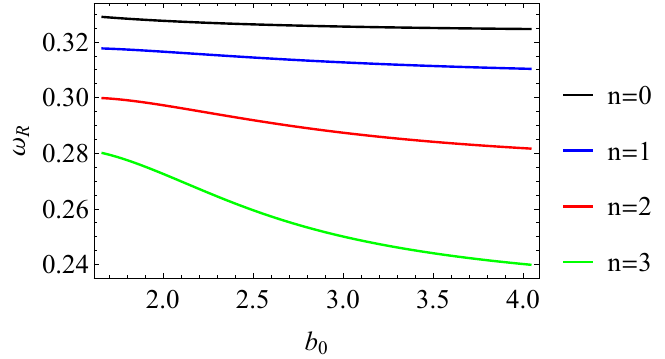}
    \end{minipage}
    \hfill
    \begin{minipage}{0.48\textwidth}
        \centering
        \includegraphics[width=\linewidth]{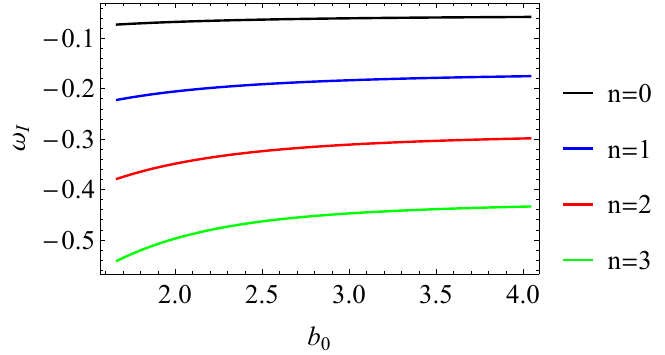}
    \end{minipage}
    \caption{For massless scalar field, QNFs of fundamental mode and the first three overtones with $l=1$ are presented as functions of the quantum parameter $b_0$ for $Q=\frac{4}{5}$. The left and right panels show the real and imaginary parts, respectively.
    }
    \label{fig:Q4c5 l1n0123 Re Im}
\end{figure}

Subsequently, we turn on the multipole quantum number $l$ and analyze the properties of the QNFs. 
Fig.\ref{fig:Q2c5 l1n012 Re Im} presents the QNFs with $l=1$ as a function of $b_0$ for $Q=\frac{2}{5}$. For the fundamental mode ($n=0$), increasing $b_0$ results in a monotonic decrease in $\omega_R$ and a monotonic increase in $\omega_I$, consistent with the $l=0$ case. For the first two overtones, $\omega_R$ exhibits non-monotonic behavior as a function of $b_0$, which is a slower outburst compared to $l=0$. This is likely due to the activation of the centrifugal potential for non-zero $l$ \cite{Zinhailo:2024kbq,Konoplya:2022hll,Bolokhov:2023bwm}, which suppresses the gravitational potential and, consequently, the quantum gravity effect. As discussed in the introduction and previous works \cite{Gong:2023ghh,Zhang:2024nny,Zinhailo:2024kbq,Konoplya:2022hll,Bolokhov:2023bwm}, this non-monotonic behavior in $\omega_R$ is a manifestation of the quantum gravity effect. When the charge parameter $Q$ is increased (see Fig.\ref{fig:Q4c5 l1n0123 Re Im} for $Q=\frac{4}{5}$), the non-monotonic behavior in $\omega_R$ disappears across the allowed $b_0$-rang. We expect that the non-monotonic behaviors and overtone outburst will reemerge at higher overtones, as observed in previous studies \cite{Zinhailo:2024kbq,Konoplya:2022hll,Bolokhov:2023bwm,Fu:2022cul,Gong:2023ghh,Zhang:2024nny,Dong:2024ams}. However, the numerical computation of higher overtone modes becomes more challenging in the presence of charge, so we defer this to future work.

\begin{figure}[H]
    \centering
    \includegraphics[width=0.5\linewidth]{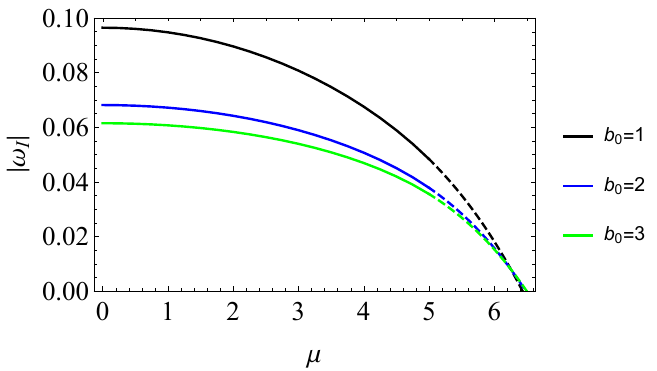}
    \caption{For massive scalar field, the imaginary parts of the frequencies as function of $\omega (\mu)$, with $l=20$, $ Q=\frac{1}{5}$, $n=0$, $b_0 =1,2,3$, and $\mu=0...6$.}
    \label{fig:quasi_resonances behavior}
\end{figure}

Finally, we activate the mass term to investigate its impact on the QNFs. Given our exclusive focus on the fundamental modes of the massive scalar field, the $6^{th}$-order WKB approximation proves sufficient for calculating the QNFs. In addition, we set $l=20$ here, where the WKB approximation can achieve high accuracy. Fig.\ref{fig:quasi_resonances behavior} displays the absolute value of the imaginary parts of the QNFs $|\omega_I|$ as a function of the scalar field mass $\mu$, with fixed charge $Q=\frac{1}{5}$, across varying values of the LQG-correction parameter $b_0$. A key observation is the smooth monotonic decrease of $|\omega_I|$ with increasing $\mu$, asymptotically approaching zero. This implies that the decay rate of the modes diminishes as the scalar field mass grows, illustrating the potential for the emergence of arbitrarily long-lived modes. Consequently, massive modes exhibit slower scattering compared to massless modes, and under specific conditions, the oscillations of massive fields become undamped—a phenomenon termed quasi-resonances \cite{Ohashi:2004wr, Konoplya:2004wg, Konoplya:2006br}. This behavior is consistent with observations in the neutral ABBV BH \cite{Bolokhov:2023bwm}. Notably, quasi-resonances have also been reported in Schwarzschild-like brane-localized black hole backgrounds \cite{Zinhailo:2024jzt}, and massive fields with non-zero spin \cite{Konoplya:2005hr, Konoplya:2017tvu}. Importantly, the existence of quasi-resonances is not universal across all spacetime geometries. For instance, they are absent in Schwarzschild-de Sitter spacetimes \cite{Konoplya:2004wg}.

\subsection{Dirac field case}\label{subsec-Dirac}

In this subsection, we analyze the QNMs of the Dirac field in the charged ABBV BH spacetime using the PSM. As mentioned in the introduction, the study of the overtone modes of the Dirac field is still absent even in the neutral ABBV BH. As a cross-validation, we also employ $6^{th}$ order WKB approximation to work out the fundamental modes (see Appendix \ref{App-WKB} for error analysis). In addition, in this subsection, we restrict our focus to the lowest admissible multipole quantum number for spin-$\frac{1}{2}$, i.e., $k=1$.

\begin{figure}[H]
    \centering
    \begin{minipage}{0.48\textwidth}
        \centering
        \includegraphics[width=\linewidth]{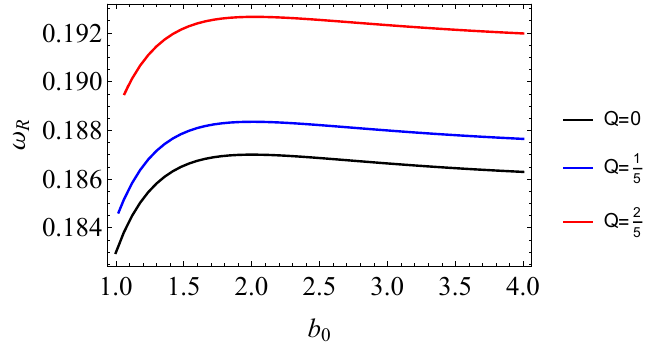}
    \end{minipage}
    \hfill
    \begin{minipage}{0.48\textwidth}
        \centering
        \includegraphics[width=\linewidth]{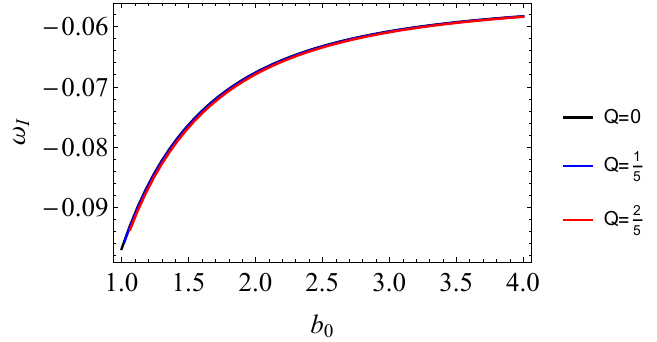}
    \end{minipage}
    \caption{For Dirac field, QNFs of fundamental modes are presented as a function of the quantum parameter $b_0$ for $k=1$, while considering various vales of the charge, specifically $Q=0, \frac{1}{5}, \frac{2}{5}$.}
    \label{fig:FFvsb0-Dirac}
\end{figure}

Fig.\ref{fig:FFvsb0-Dirac} illustrates the QNFs of the Dirac field as a function of the LQG-corrected parameter $b_0$, with fixed charge $Q$. Notably, the real part of the QNFs $\omega_R$ manifests a pronounced non-monotonic dependence on $b_0$, characterized by an initial increase followed by a gradual decline. This pattern is qualitatively similar to the $\omega_R$ behavior in the neutral ABBV BH spacetime (the black curve for $Q=0$ and also see Ref. \cite{Bolokhov:2023bwm}). In contrast, the scalar field case displays monotonically decreasing behavior across the same $b_0$ range ($l=0$; see left panel of Fig.\ref{fig:FFvsb0}). Meanwhile, the imaginary part $\omega_I$ demonstrates monotonically increasing magnitudes with $b_0$, exhibiting significantly greater variation compared to the RN BH baseline. This amplification of $\omega_I$'s sensitivity to LQG corrections aligns with the scalar field behavior (right panel of Fig. \ref{fig:FFvsb0}), suggesting a potential universality in how quantum-gravity-induced modifications diminish damping rates across scalar and Dirac perturbations.

\begin{figure}[H]
    \centering
    \begin{minipage}{0.45\textwidth}
        \centering
        \includegraphics[width=\linewidth]{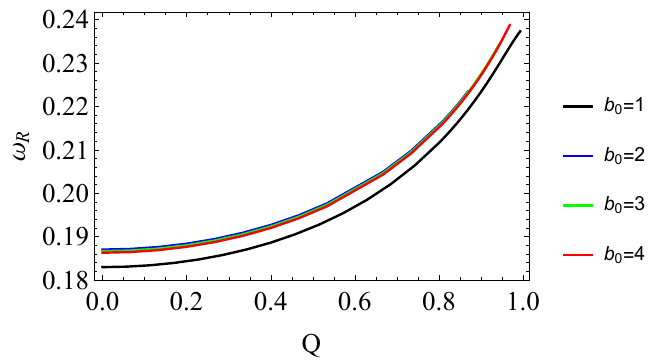}
    \end{minipage}
    \hfill
    \begin{minipage}{0.45\textwidth}
        \centering
        \includegraphics[width=\linewidth]{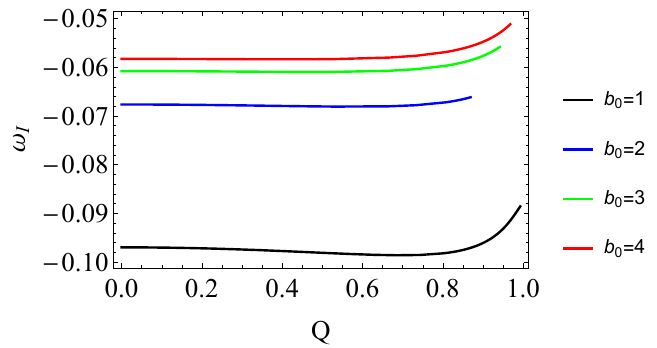}
    \end{minipage}
    \caption{For Dirac field, QNFs of fundamental modes are presented as a function of the charge $Q$ for $k=1$, while considering various values of the quantum parameter, specifically $b_0=1,2,3,4$.}
    \label{fig:FFvsQ-Dirac}
\end{figure}

Fig.\ref{fig:FFvsQ-Dirac} depicts the QNFs of the Dirac field as functions of the charge $Q$. The real part of the QNFs $\omega_R$ displays a strict monotonic increase with $Q$, while the imaginary part $\omega_I$ remains approximately constant across the majority of the $Q$-range, deviating only through a subtle upward trend at large $Q$. This qualitative agreement with the scalar field's QNF behavior under LQG corrections (right panel of Fig. \ref{fig:FFvsQ}) suggests a universal response of scalar and Dirac perturbations to charge-dependent quantum spacetime modifications. Notably, even in the RN limit ($b_0=1$), the dependence of the $\omega_R$ on $Q$ for Dirac field is strictly monotonic (black curve in the left panel of Fig.\ref{fig:FFvsQ-Dirac}). This behavior stands in sharp contrast to the scalar field's non-monotonic $Q$-dependence observed under identical RN limit (see black curve for $b_0=1$ in the left panel of Fig.\ref{fig:FFvsQ}).

\begin{figure}[H]
    \centering
    \begin{minipage}{0.48\textwidth}
        \centering
        \includegraphics[width=\linewidth]{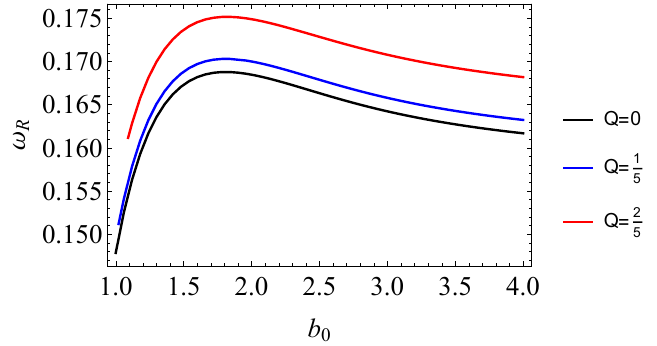}
    \end{minipage}
    \hfill
    \begin{minipage}{0.48\textwidth}
        \centering
        \includegraphics[width=\linewidth]{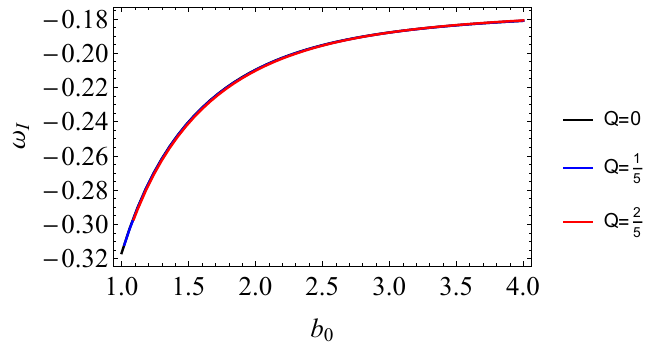}
    \end{minipage}
    \caption{For Dirac field, QNFs of the first overtone ($n=1$) with $k=1$ are presented as a function of the quantum parameter $b_0$ for the values of the charge $Q=0, \frac{1}{5}, \frac{2}{5}$. The left and right panels show the real and imaginary parts, respectively.}
    \label{fig:omegaRvsb0n1-Dirac}
\end{figure}
\begin{figure}[H]
    \centering
    \begin{minipage}{0.48\textwidth}
        \centering
        \includegraphics[width=\linewidth]{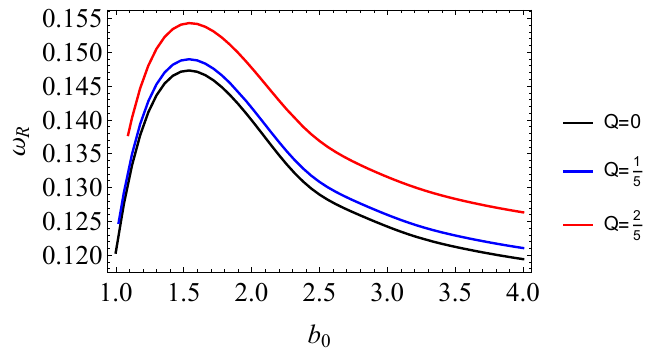}
    \end{minipage}
    \hfill
    \begin{minipage}{0.48\textwidth}
        \centering
        \includegraphics[width=\linewidth]{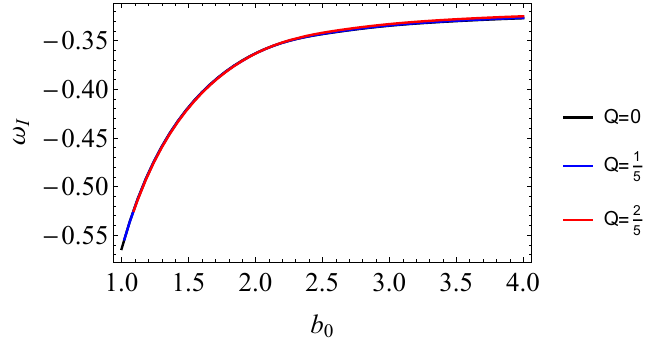}
    \end{minipage}
    \caption{For Dirac field, QNFs of the second overtone ($n=2$) with $k=1$ are presented as a function of the quantum parameter $b_0$ for the values of the charge $Q=0, \frac{1}{5}, \frac{2}{5}$. The left and right panels show the real and imaginary parts, respectively.}
    \label{fig:omegaRvsb0n2-Dirac}
\end{figure}

We now investigate the overtone spectra of the Dirac field. Fig.\ref{fig:omegaRvsb0n1-Dirac} ($n=1$) and Fig.\ref{fig:omegaRvsb0n2-Dirac} ($n=2$) display the QNFs as functions of the quantum parameter $b_0$ for varying charge $Q$. A key observation is the enhanced non-monotonicity of $\omega_R$ in overtones compared to fundamental modes ($n=0$), although no pronounced outbursts or oscillatory patterns emerge in the first two overtones. This contrasts with the scalar field dynamics for the lowest admissible multipole quantum number ($l=0$), where quantum gravity effects induce both overtone outbursts and oscillatory behavior. However, we note that for the scalar field with $l=1$, these outbursts and oscillatory patterns are also replaced by non-monotonic behavior. We hypothesize that similar outbursts or oscillatory patterns may reemerge in the Dirac field at higher overtones, a possibility we reserve for future investigation. Notably, the imaginary part $\omega_I$ retains its monotonic growth with $b_0$, consistent with the behavior observed in fundamental modes.
\begin{figure}[H]
    \centering
    \begin{minipage}{0.48\textwidth}
        \centering
        \includegraphics[width=\linewidth]{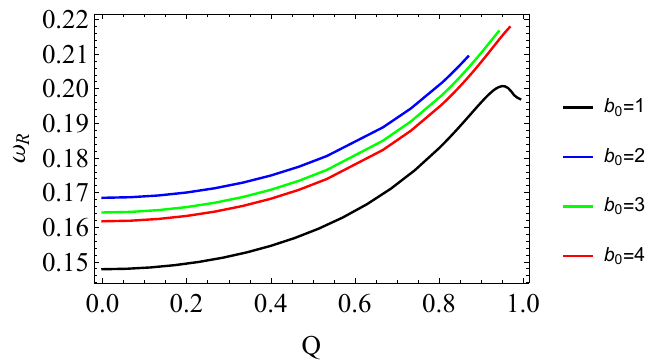}
    \end{minipage}
    \hfill
    \begin{minipage}{0.48\textwidth}
        \centering
        \includegraphics[width=\linewidth]{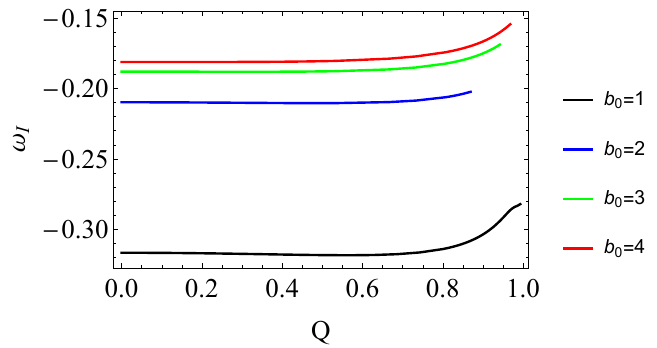}
    \end{minipage}
    \caption{For Dirac field, QNFs of the first overtone ($n=1$) with $k=1$ are presented as a function of the charge $Q$ for different quantum parameters $b_0=1,2,3,4$. The left and right panels show the real and imaginary parts, respectively.}
    \label{fig:omegaRvsQn1-Dirac}
\end{figure}
\begin{figure}[H]
    \centering
    \begin{minipage}{0.48\textwidth}
        \centering
        \includegraphics[width=\linewidth]{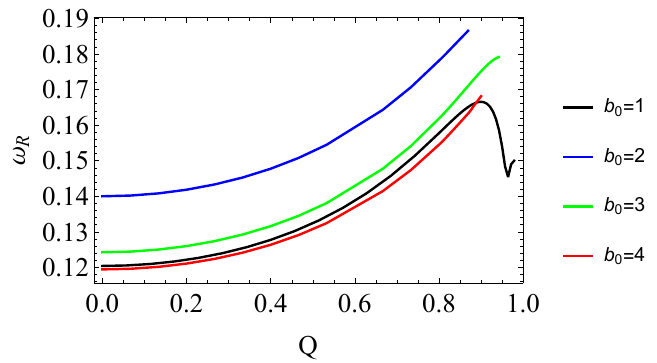}
    \end{minipage}
    \hfill
    \begin{minipage}{0.48\textwidth}
        \centering
        \includegraphics[width=\linewidth]{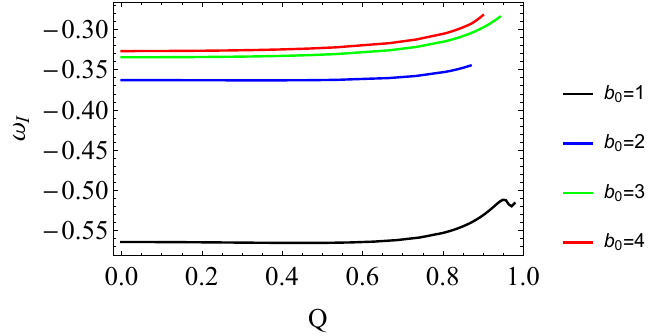}
    \end{minipage}
    \caption{For Dirac field, QNFs of the second overtone ($n=2$) with $k=1$ are presented as a function of the charge $Q$ for different quantum parameters $b_0=1,2,3,4$. The left and right panels show the real and imaginary parts, respectively.}
    \label{fig:omegaRvsQn2-Dirac}
\end{figure}

Fig.\ref{fig:omegaRvsQn1-Dirac} ($n=1$) and Fig.\ref{fig:omegaRvsQn2-Dirac} ($n=2$) exhibit the QNFs as functions of charge $Q$ under fixed quantum parameter $b_0$. In the RN BH limit ($b_0=1$), we observe a clear non-monotonic dependence of $\omega_R$ on $Q$ for $n=1$ as $Q\rightarrow 1$. This non-monotonicity transitions into a weak amplitude outburst and oscillatory pattern for $n=2$. But the outburst and oscillatory is weaker than their scalar field counterparts with $l=0$ (cf. Subsection \ref{subsec-Scalar}). Remarkably, when quantum corrections are activated ($b_0>1$), all anomalous features – non-monotonicity, outbursts, and oscillations – are entirely suppressed for the first two overtones across the allowed $Q$-range. This suppression effect aligns with our earlier findings in the scalar field case, suggesting a universal interplay between charge $Q$ and quantum parameter $b_0$.

\section{Conclusions and discussions}
\label{sec:conclusion}

In this work, we have systematically analyzed the properties of the QNM spectra of massless/massive scalar and Dirac fields around a charged ABBV BH, unveiling novel signatures of quantum gravity effects and their interplay with the charge. Our key findings and their implications are summarized as follows:
\begin{itemize}
    \item The LQG-corrected parameter $b_0$ triggers a distinct overtone outburst in the QNM specta, particularly pronounced for the scalar field with $l=0$. This outburst, usually accompanying with an oscillatory pattern, grow more pronounced with increasing overtone numbers, marking a direct imprint of quantum gravity effect on the BH spectroscopy. For the scalar field with $l>0$ and the Dirac field, however, the overtone outburst develops more slowly or degenerates into non-monotonic on $b_0$, suggesting a competition between quantum corrections and centrifugal potential barriers. These anomalies provide a unique spectral fingerprint to observationally discriminate LQG-BHs from their classical counterparts.
    \item Increasing the charge parameter $Q$ universally suppresses quantum-gravity-induced spectral features—outbursts, non-monotonicity, and oscillations. 
    \item Our findings provide evidence for the existence of quasi-resonances in the massive scalar QNM spectrum, highlighting the potential for the occurrence of arbitrarily long-lived modes in this charged LQG spacetime.
\end{itemize}

Future studies should extend this work to rotating LQG-BHs (e.g., quantum Kerr analogs), where frame-dragging and ergoregion instabilities may amplify quantum spectral features. Further, a deeper understanding of the physical mechanism behind the overtone outburst and the role of different parameters in governing this phenomenon could shed light on new physics beyond GR.

\acknowledgments

We would like to express our sincere gratitude to the anonymous reviewer for the insightful and valuable suggestions. We are especially grateful to Prof. Rui-Hong Yue for helpful discussions and suggestions.
This work is supported by National Key R$\&$D Program of China (No. 2020YFC2201400), the Natural Science Foundation of China under Grants No. 12375055, 12347159 and 12447151.

\newpage
\appendix
\section{Error analysis}\label{App-WKB}

To establish the numerical fidelity of our PSM-based QNM calculations, we perform a systematic cross-validation against the WKB approximation, focusing on fundamental modes ($n=0$) and low overtones ($n\leq 2$).

The WKB method — a semi-analytical approach requiring only the maximum value of the effective potential $V_{\text{eff}}(r_*)$  and its derivatives — provides an efficient QNF calculating method for $n<l$. The general WKB formula could be written as:
\begin{align}
       {\omega ^2} = &{V_0} + {A_2}({{\cal K}^2}) + {A_4}({{\cal K}^2}) + {A_6}({{\cal K}^2}) +  \ldots 
            \nonumber
            \\
        &- i{\cal K}\sqrt { - 2{V_2}} (1 + {A_3}({{\cal K}^2}) + {A_5}({{\cal K}^2}) + {A_7}({{\cal K}^2}) +  \ldots )\,,
\end{align}
where ${\cal K}=n+1/2$ is half-integer. The correction terms ${A_k}({{\cal K}^2})$ are polynomials of ${\cal K}^2$ whose rational coefficients depend on higher-order derivatives of $V_{\text{eff}}(r_*)$ at its maximum. For computational robustness, we employ the $13^{th}$ WKB with Pad\'e approximation \cite{Konoplya:2019hlu} for massless scalar fields and the sixth-order WKB approximation to calculate the QNFs, respectively.

\begin{figure}[H]
    \centering
    \begin{minipage}{0.48\textwidth}
        \centering
        \includegraphics[width=\linewidth]{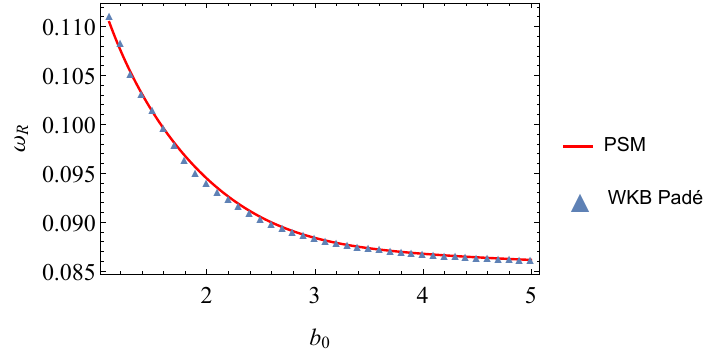}
    \end{minipage}
    \hfill
    \begin{minipage}{0.48\textwidth}
        \centering
        \includegraphics[width=\linewidth]{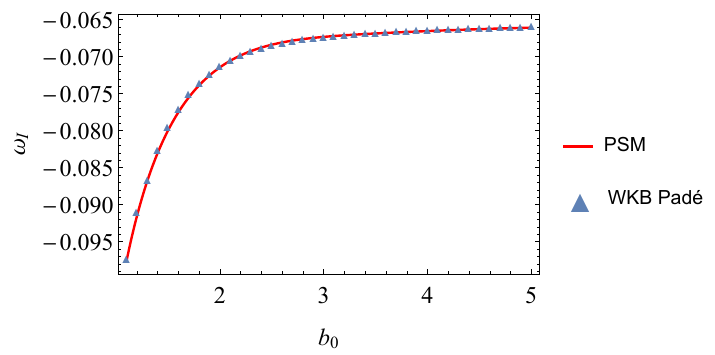}
    \end{minipage}
    \caption{QNFs of the massless scalar field as functions of the quantum parameter $b_0$, computed via PSM and  $13^{th}$ WKB approximation with Pad\'e approximation. Here, $Q=\frac{2}{5},l=0, n=0$.}
    \label{fig:Error_l0n0_Re_Im PSM WKB}
\end{figure}

\begin{figure}[H]
    \centering
    \begin{minipage}{0.48\textwidth}
        \centering
        \includegraphics[width=\linewidth]{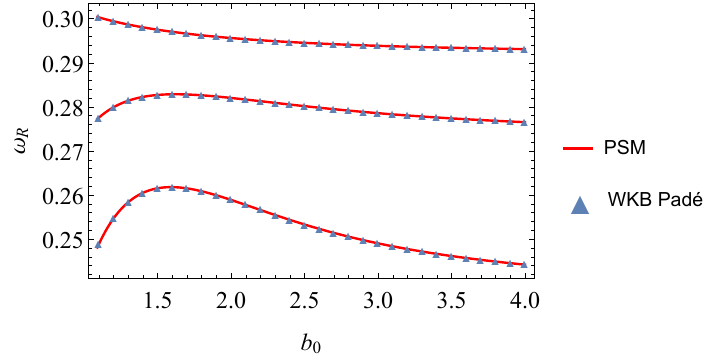}
    \end{minipage}
    \hfill
    \begin{minipage}{0.48\textwidth}
        \centering
        \includegraphics[width=\linewidth]{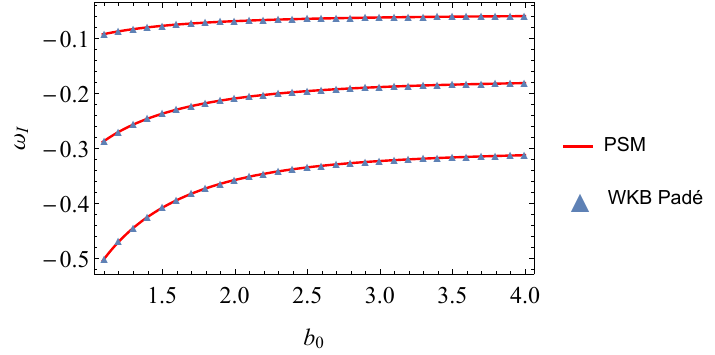}
    \end{minipage}
    \caption{QNFs of the massless scalar field as functions of the quantum parameter $b_0$, computed via PSM and  $13^{th}$ WKB approximation with Pad\'e approximation. Here, $Q=\frac{2}{5},l=1$ and $n=0,1,2$ (from top to bottom).}
    \label{fig:Error_l1n012_Re_Im PSM WKB}
\end{figure}

\begin{figure}[H]
    \centering
    \begin{minipage}{0.48\textwidth}
        \centering
        \includegraphics[width=\linewidth]{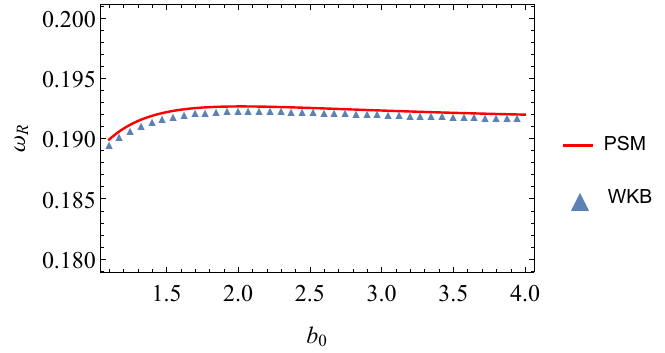}
    \end{minipage}
    \hfill
    \begin{minipage}{0.48\textwidth}
        \centering
        \includegraphics[width=\linewidth]{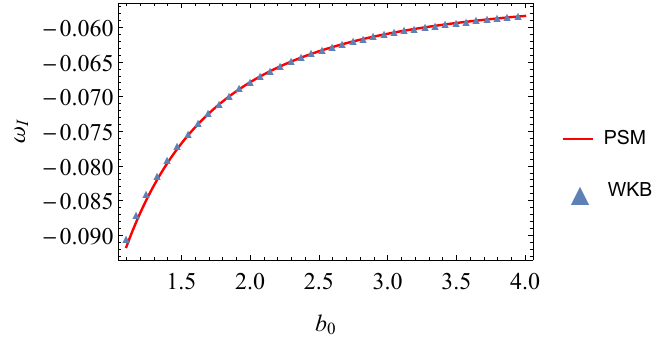}
    \end{minipage}
    \caption{QNFs of the Dirac field as functions of the quantum parameter $b_0$, computed via PSM and sixth-order WKB approximation. Here, $Q=\frac{2}{5},k=1, n=0$.}
    \label{fig:Dirac_Error_l1n0_Re_Im PSM WKB}
\end{figure}

Fig.\ref{fig:Error_l0n0_Re_Im PSM WKB}, Fig.\ref{fig:Error_l1n012_Re_Im PSM WKB} and Fig.\ref{fig:Dirac_Error_l1n0_Re_Im PSM WKB} show PSM-computed QNFs against WKB predictions for massless scalar field with $l=0, n=0$ (Fig.\ref{fig:Error_l0n0_Re_Im PSM WKB}) and $l=1, n=0,1,2$ (Fig.\ref{fig:Error_l1n012_Re_Im PSM WKB}), and Dirac field with $l=1, n=0$ (Fig.\ref{fig:Dirac_Error_l1n0_Re_Im PSM WKB}). Qualitatively, the spectral overlap between PSM and WKB is striking across all cases. To quantify this concordance, we define the relative percentage deviations:
\begin{equation}
    \varepsilon_{Re} = \left| \frac{\omega_{PS}^{Re} - \omega_{WKB}^{Re}}{\omega_{PS}^{Re}} \right| \times 100\% \,,\,\,\,\,\,
    \varepsilon_{Im} = \left| \frac{\omega_{PS}^{Im} - \omega_{WKB}^{Im}}{\omega_{PS}^{Im}} \right| \times 100\%\,.
\end{equation}
Here, $\varepsilon_{Re}$ and $\varepsilon_{Im}$ denote the percentage deviations in the real and imaginary parts of the QNFs, respectively. These metrics are computed relative to the PSM results, which serve as the reference standard for this comparison.

\begin{table}[bthp]
    \centering
    \renewcommand{\arraystretch}{1.2}  
    \begin{tabular}{p{0.7cm} p{0.7cm} p{4.5cm} p{4.5cm} p{2cm} p{2cm}}
        \hline
        $Q$ & $b_0$ & $\omega_{\text{PSM}}$ & $\omega_{\text{WKB Pad\'e}}$ & $\varepsilon_{Re}$ (\%) & $\varepsilon_{Im}$ (\%) \\
        \hline
        $2/5$ & 1.1 &0.110431 - 0.0976727$i$  &0.111010 - 0.0973977  $i$&0.5243   &0.2815 \\
              & 1.5 &0.101271 - 0.0800035$i$  &0.101383 - 0.0795293  $i$&0.1105   &0.5927 \\
              & 2   &0.0944658 - 0.0714582$i$ &0.0939013 - 0.0713471 $i$&0.5976   &0.1556 \\
              & 2.5 &0.0905033 - 0.0683967$i$ &0.0903154 - 0.0684605 $i$&0.2076   &0.0933 \\
              & 3   &0.0883746 - 0.0673379$i$ &0.0883085 - 0.0673566 $i$&0.0748  &0.0279 \\
              & 3.5 &0.0873154 - 0.0668779$i$ &0.0872944 - 0.0667929 $i$&0.0241  &0.1271 \\
              & 4   &0.0867482 - 0.0665548$i$ &0.0866981 - 0.0664276 $i$&0.0578  &0.1912 \\
        \bottomrule
    \end{tabular}
    \caption{Comparison of the relative errors $\varepsilon_{Re/Im}$ in the QNFs for a massless scalar field, computed via PSM and $13^{th}$ WKB method with Pad\'e approximation, for various $b_0$ with $Q=\frac{2}{5}$, $l=0$ and $ n=0$.}
    \label{tab:error for massless scalar l0 n0 WKBpade PSM}
\end{table}
\begin{table}[bthp]
    \centering
    \renewcommand{\arraystretch}{1.2}  
    \begin{tabular}{p{0.7cm} p{0.7cm} p{4.5cm} p{4.5cm} p{2cm} p{2cm}}
        \hline
        $Q$ & $b_0$ & $\omega_{\text{PSM}}$ & $\omega_{\text{WKB Pad\'e}}$ & $\varepsilon_{Re}$ (\%) & $\varepsilon_{Im}$ (\%) \\
        \hline
        $2/5$ & 1.1 &0.300270 - 0.0922177$i$  & 0.300270 - 0.0922177$i$& 0  &0 \\
              & 1.5 &0.297472 - 0.0772634$i$  & 0.297472 - 0.0772634$i$& 0  &0 \\
              & 2   &0.295583 - 0.0686968$i$  & 0.295583 - 0.0686968$i$& 0  &0 \\
              & 2.5 &0.294495 - 0.0644370$i$  & 0.294495 - 0.0644371$i$&  0 & 0\\
              & 3   &0.293819 - 0.0620358$i$  & 0.293819 - 0.0620358$i$&  0 & 0\\
              & 3.5 &0.293374 - 0.0605580$i$  & 0.293374 - 0.0605580$i$& 0  &0 \\
              & 4   &0.293069 - 0.0595871$i$  & 0.293069 - 0.0595871$i$&  0 & 0\\
        \bottomrule
    \end{tabular}
    \caption{Comparison of the relative errors $\varepsilon_{Re/Im}$ in the QNFs for a massless scalar field, computed via PSM and $13^{th}$ WKB method with Pad\'e approximation, for various $b_0$ with $Q=\frac{2}{5}$, $l=1$ and $ n=0$. An error of $0\%$ indicates that the difference is smaller than $10^{-4}\%$.}
    \label{tab:error for massless scalar l1 n0 WKBpade PSM}
\end{table} 
\begin{table}[bthp]
    \centering
    \renewcommand{\arraystretch}{1.2}  
    \begin{tabular}{p{0.7cm} p{0.7cm} p{4.5cm} p{4.5cm} p{2cm} p{2cm}}
        \hline
        $Q$ & $b_0$ & $\omega_{\text{PSM}}$ & $\omega_{\text{WKB}}$ & $\varepsilon_{Re}$ (\%) & $\varepsilon_{Im}$ (\%) \\
        \hline
        $2/5$ & 1.1 &0.189906 - 0.0917185$i$  & 0.189393 - 0.0905825$i$& 0.2705  &1.2386 \\
              & 1.5 &0.192179 - 0.0768029$i$  & 0.191661 - 0.0766089$i$& 0.2694  &0.2526 \\
              & 2   &0.192652 - 0.0679596$i$  & 0.192230 - 0.0679397$i$& 0.2194  &0.0294 \\
              & 2.5 &0.192527 - 0.0634864$i$  & 0.192144 - 0.0634900$i$& 0.1992  &0.0056\\
              & 3   &0.192314 - 0.0609546$i$  & 0.191948 - 0.0609694$i$& 0.1899  &0.0242\\
              & 3.5 &0.192123 - 0.0593985$i$  & 0.191769 - 0.0594200$i$& 0.1841  &0.0361 \\
              & 4   &0.191971 - 0.0583792$i$  & 0.191625 - 0.0584046$i$& 0.1804  &0.0435\\
        \bottomrule
    \end{tabular}
    \caption{Comparison of the relative errors $\varepsilon_{Re/Im}$ in the QNFs for a Dirac field, computed via PSM and sixth-order WKB method, for various $b_0$ with $Q=\frac{2}{5}$, $k=1$ and $ n=0$.}
    \label{tab:error for Dirac scalar l1 n0 WKBpade PSM}
\end{table} 

Tables \ref{tab:error for massless scalar l0 n0 WKBpade PSM}–\ref{tab:error for Dirac scalar l1 n0 WKBpade PSM} detail the deviations $\varepsilon_{Re/Im}$ for fixed $Q$ and various values of $b_0$. Key findings:
\begin{itemize}
    \item \textbf{Massless scalar field with $l=0, n=0$:} $\varepsilon_{Re/Im}\lesssim 10^{-1}\%$,
    \item \textbf{Massless scalar field with $l=1, n=0$:} $\varepsilon_{Re/Im}<10^{-4}\%$,
    \item \textbf{Dirac field with $k=1, n=0$:} $\varepsilon_{Re/Im}\lesssim 1\%$.
\end{itemize}
The strong consistency between the WKB and PSM results underscores the reliability of our PSM-based QNM calculations. This agreement serves as a critical cross-verification, highlighting the robustness of PSM-based QNM analysis in this work.

\section{The reduced wave equations}
\label{Appendix:equation for PSM}

By imposing the appropriate boundary conditions, the wave equation \eqref{drast-dr} reduces to the following form:
\begin{equation}
    {\alpha _0}\psi '' + {\beta _0}\psi ' + {\gamma _0}\psi  = 0\,,
\end{equation}
where the coefficients are determined by the specific field under consideration.
For the massless scalar field, the coefficients are given by:
\begin{align}
{\alpha_0}=&\,2b_0^4{u^2}f(u)\left( {m - {r_0}g(u)} \right)\,,
\
\\
\nonumber
{\beta _0}=&\,ib_0^2{u^2}f(u)\left[ -8{\left(m-{r_0}\omega g(u) \right)\left( { - mu + b_0^2{r_h} + 3b_0^2mu} \right)\omega + ib_0^2{r_0}{u^2}g'(u)} \right]
\nonumber
\
\\
&
+ 2b_0^4{u^2}\left[ {2i\sqrt m {r_h}\omega \sqrt {m - {r_0}g(u)}  + {u^2}\left( {m - {r_0}g(u)} \right)f'(u)} \right]\,,
 \
 \\
{\gamma _0}=&\,2i{b_0}^2{u^2}\left[ {- mu + b_0^2\left( {{r_h} + 3mu} \right)} \right]\omega \left( {m - {r_0}g(u)} \right)f'(u) - ib_0^2{r_0}{u^2}\left( { - mu + b_0^2{r_h} + 3b_0^2mu} \right)g'(u)
\nonumber
\
\\
&
+ b_0^2\sqrt m \left\{ {b_0^2\left({l^2}+l\right)\sqrt m {u^2}+\left[{4m{r_h}u{\omega ^2} - 2{r_h}\omega \left( - iu + 2{r_h}\omega  + 6mu\omega\right)}\right]
\sqrt {m - {r_0}g(u)} } \right\}
\nonumber
\
\\
&
+\left\{ {2{m^2}u\left[ {2mu\omega  + b_0^2\left( { - 4{r_h}\omega  + iu - 12mu\omega } \right)} \right] + b_0^4 \left[2r_h^2\omega  - u\left( {2{r_h} + 3mu} \right)\left( {i - 6m\omega } \right)\right]} \right\}
\nonumber
\
\\
&
\left( {\omega f(u) - 2{r_0}g\left( u \right)} \right)\,.
\end{align}
For the Dirac field, the coefficients are:
\begin{align}
{\alpha_0}=\,& - 2u{\left( {1 - {u^2}} \right)^{4}}f{\left( u \right)^2}b_0^4\left( {m - g\left( u \right){r_0}} \right)\,, \\
{\beta _0}=\,& \left\{ b_0^2{{\left( {{u^2} - 1} \right)}^2}\left[ { - 2uf'(u)\left( {m - {r_0}g(u)} \right) + f(u){r_0}\left( {ug'(u) - 2g(u)} \right) + 2f(u)m} \right] \vphantom{\sqrt {1 - \frac{{{r_0}g(u)}}{m}}} \right.
\nonumber \\
&\, \left. + 8imb_0^2{u^2}\omega {r_h}\sqrt {1 - \frac{{{r_0}g(u)}}{m}}  + 16i{u^2}\omega f(u)\left[ { - b_0^2\left( {{r_h} - 3{u^2} + 3} \right) - {u^2} + 1} \right]\left( {m - {r_0}g(u)} \right) \right\}
\nonumber \\
&\, \times b_0^2f(u){\left( {1 - {u^2}} \right)^{2}}\,, \\
{\gamma _0}=&\, \left\{ {4{u^2}\omega {{\left( {b_0^2\left( {{r_h} - 3{u^2} + 3} \right) + {u^2} - 1} \right)}^2} + ib_0^2\left( {{u^2} - 1} \right)\left\{ {b_0^2\left[ {\left( {3{u^2} + 1} \right){r_h} - 3{u^4} + 3} \right] + {u^4} - 1} \right\}} \right\}
\nonumber \\
&\, \times 8u\omega f{(u)^2}\left( {m - {r_0}g(u)} \right) 
\nonumber \\
&\, +\left\{ f(u)\left\{ { - 2\left( {{u^2} - 1} \right)\left[ {\left( {{u^2} - 1} \right)f'(u)\left( {m - {r_0}g(u)} \right) + 2{k^2}mu} \right] + 8imu\omega {r_h}\sqrt {1 - \frac{{{r_0}g(u)}}{m}} } \right\} + \right.
\nonumber \\
&\, \left. km{{\left( {{u^2} - 1} \right)}^2}\sqrt {f(u)} f'(u)\sqrt {1 - \frac{{{r_0}g(u)}}{m}}  + r{{\left( {{u^2} - 1} \right)}^2}\left[ {{r_0}f{{(u)}^2}g'(u) + 4kmuf{{(u)}^{3/2}}\sqrt {1 - \frac{{{r_0}g(u)}}{m}} } \right] \right\}
\nonumber \\
&\, \times 2b_0^4{u^2}{\left( {1 - {u^2}} \right)}
\nonumber \\
&\, +\left\{  - 2{{\left( {{u^2} - 1} \right)}^2}uf'(u)\left( {m - {r_0}g(u)} \right) + {{\left( {{u^2} - 1} \right)}^2}f(u)\left[ {{r_0}\left( {ug'(u) - 2g(u)} \right) + 2m} \right]  \vphantom{\sqrt {1 - \frac{{{r_0}g(u)}}{m}}}
\right.
\nonumber \\
&\, \left. + 8im{u^2}\omega {r_h}\sqrt {1 - \frac{{{r_0}g(u)}}{m}}  \right\} \times \left( {b_0^2\left( {{r_h} - 3{u^2} + 3} \right) + {u^2} - 1} \right)
\nonumber \\
& \, \times 4ib_0^2u\omega f(u)\,.
\end{align}

\bibliographystyle{style1}
\bibliography{Ref}
 \end{document}